\documentclass[12pt,journal,oneside,onecolumn]{IEEEtran}
\pdfoutput=1
\newcounter{mytempeqncnt}
\usepackage{amsmath, amssymb,amscd, epsfig}

\usepackage{graphicx}
\usepackage{mathrsfs}
\usepackage{cite}
\usepackage{algorithm}
\usepackage{algorithmic}
\usepackage{color}
\usepackage{multirow}

\begin{document}
\title{Delay Analysis of Multichannel Opportunistic Spectrum Access MAC Protocols}
\author{Arash Azarfar,~\IEEEmembership{Student Member,~IEEE}, Jean-Fran\c{c}ois Frigon,~\IEEEmembership{Senior Member,~IEEE}, and Brunilde Sans\`o,~\IEEEmembership{Member,~IEEE}%
\thanks{The authors are from the Department of Electrical Engineering, \'{E}cole Polytechnique de Montr\'{e}al, Montr\'{e}al (Qc), Canada, H3T 1J4 (e-mail: \{arash.azarfar, j-f.frigon, brunilde.sanso\}@polymtl.ca). \newline \indent This research project was supported by NSERC under Grant STPG365205. \newline This work has been submitted to the IEEE for possible publication. Copyright may be transferred without notice, after which
this version may no longer be accessible.}
}
%

\maketitle
\begin{abstract}
We provide in this paper a comprehensive delay and queueing analysis for two baseline 
medium access control (MAC) protocols for  
multi-user cognitive radio (CR) networks and investigate the impact 
of different network parameters, such as packet size, Aloha-type medium access 
probability and number of channels on the system performance. 
In addition to an accurate Markov chain, which follows the queue status of all users, 
several lower complexity queueing theory approximations are provided. Accuracy and performance 
of the proposed analytical approximations are verified with extensive simulations. It is observed 
that for CR networks using an Aloha-type access to the control channel, a buffering MAC protocol, 
where in case of interruption the CR user waits for the primary user to vacate the channel before resuming the
transmission, outperforms 
a switching MAC protocol, where the CR user vacates the channel in case of appearance of 
primary users and then compete again to gain access to a new channel. The reason is that the delay
bottleneck for both protocols is the time required to successfully access the control channel,
which occurs more frequently for the switching MAC protocol. We also propose a user clustering approach, 
where users are divided into clusters with a separate control channel per 
cluster, and observe that it can significantly improve the performance by reducing the number 
of competing users per control channel. 
\end{abstract}

\section{Introduction}
\label {sec:intro}

Opportunistic spectrum access (OSA) communication models \cite{zhao07a}, implemented by cognitive radios (CR), offer both the capacity to decrease the communication infrastructure expenses by using the vacant portions of the spectrum and to improve the quality of wireless communications by permitting wireless CR nodes\footnote{\emph{Node} and \emph{user} are used interchangeably.} to switch to a better channel when the channel quality is not satisfactory. These notable features have made cognitive-radio based wireless networks a technology of choice for incoming wireless technologies.
Frequent spectrum handovers along with new requirements such as spectrum sensing and spectrum decision distinguish a cognitive radio network from its predecessors. New medium access control (MAC) protocols which take into account the inherent nature of cognitive radios are thus indispensable. For this aim, several  medium access protocols have been proposed in the literature \cite{zhao07,jia08,cordeiro07}. Different protocols may differ in details such as the existence of a common control channel, number of radios required for the nodes and sensing algorithm (please see \cite{cormio09,pawelczak09} and references therein for a detailed survey of CR MAC protocols). Since most of the proposed MAC schemes assume a dedicated control channel, and given that being equipped with multiple radios increases the hardware and complexity cost, we investigate in this paper a realistic cognitive radio network with a common control channel, distributed sensing appropriate for ad-hoc networks, and a single radio per node. 

In the literature, most of the attention on the analysis of the different medium access protocols and their variations has been focused on the throughput analysis in the presence of saturated traffic (e.g., \cite{pawelczak09,park11,zhao07,pawelczak08,jia08,cordeiro07}), while little work has been done on delay and jitter analysis. Throughput analysis provides an upper bound for the performance of the network, which can be used for evaluation and comparison purposes of different protocols. However, delay analysis and packet level performance evaluation are required to investigate how a protocol, along with its parameter settings, behaves for delay-sensitive applications, such as multimedia communications \cite{wang-l11,cisco13}.

In the few papers available on delay analysis (e.g., \cite{rashid07,S-Wang12, wang12,laourine10,kim12}), detailed system parameters are not considered or the CR network under consideration is simple, such that full insights into the impact of different parameters on the network performance have not been provided. The target has also been specific networks; e.g., sensor networks, with specific requirements \cite{bicen12}, such that the results can not be easily extended to general models. 
In \cite{azarfar12e,azarfar13f}, we proposed a general M/G/1 queueing model~\cite{kleinrock75} for a cognitive radio link where the server is subject to interruptions~\cite{takagi91}.
The channel (i.e., the server of the queue) is subject to interruptions because the CR node has to suspend its communication session for the period of spectrum handover (switching) or primary users' activity (buffering).
The proposed models in those papers  were for a single node with continuous-time operating and interruption periods. However, those models didn't address the complex problem of multi-node operation in a multi-channel environment. 

Our objective in this paper is thus to fill this MAC protocols analysis gap by providing a comprehensive delay analysis for two baseline medium access control protocols discussed in \cite{pawelczak09,park11}. Note that the models described in those papers are not detailed MAC models for practical purposes. 
The intent is rather to have models with enough details, yet general enough, to use as baseline reference models 
for families of MAC protocols and investigate the effect of different parameters on the throughput of a cognitive radio network. 
In this paper, we have the same objective for the delay analysis and thus only relax the saturated traffic 
assumption and keep the same other assumptions and modeling details to analyze the average delay. 
In the first part, a buffering MAC model~\cite{pawelczak09,park11}, in which a node stays on the channel in case of interruptions 
until the transmission of the packet is finished, is investigated. In the second part, a switching MAC model~\cite{park11}, 
where the node leaves the interrupted channel and participates in a new reservation competition every time that an interruption occurs, is investigated.

The main contribution of this work is thus providing a comprehensive analytical delay analysis for a multichannel cognitive radio network 
with both buffering and switching recovery policies where the impact of different network parameters 
(number of users, number of channels, control channel access probability, channel availability, and arrival rate) 
on the performance is considered. The results can thus be used in network optimization and design problems.  
For both MAC models, we derive an exact Markov chain model which can be solved to obtain the system time distribution and moments. Since the number of states in the Markov chains grows exponentially with the number of nodes, we also propose lower complexity approximations based on discrete-time M/G/1 queueing theory results. In both MAC models, the service time of a packet is composed of two parts: the time spent to compete with other users to reserve a channel (for the switching MAC model, the reservation period may occur multiple times during the service of a single packet), and the transmission time.  We use Markov chain models, combined with Renewal theory results, to derive approximate distributions of both parts of the packet service time. The service time moments are then used to find the approximate average system time of packets in the OSA network for both MAC protocols. We also provide numerical results to validate the analysis and provide insight on the delay performance of the MAC protocols for different network parameters.

The paper proceeds as follows.
In Section \ref{sec:system-model}, the system model is presented and related parameters are introduced. The next two sections discuss the buffering model. In Section \ref{sec:Queue-Occupancy}, an accurate Markov chain for queue occupancy is proposed from which different parameters of interest can be derived. As the queue occupancy Markov chain is not scalable, a service cycle analysis is provided in Section \ref{sec:Service-Cycle}. Section \ref{sec:switching} discusses the switching model where the same approach is followed: first an accurate Markov chain for queue occupancy is proposed and then a delay cycle analysis and related approximations are provided. Numerical and simulation results for both buffering and switching models are discussed in Section \ref{sec:simulation}. Finally Section \ref{sec:conclusion} concludes the paper with some remarks on future research direction. The notation used in this paper are listed in Table \ref{tab:notations}.

\begin{table}[]
\caption{{Notations}}
\label{tab:notations}
\centering
\begin{tabular} {c|l}
\hline
\scriptsize
{\textbf{Notation}} & {\textbf {Description}} \\
 \hline
$N$& Number of CR nodes\\
$M$& Total number of channels\\
$M_C$& Number of data channels ($=M-1$)\\
$\lambda$& Poisson packet arrival rate\\
$X$& Geom/G/1 service time \\
$X_R$& Reservation time \\
$X_T$& Transmission time (buffering) \\
$X_U$& Interrupted transmission time (switching) \\
$L$& Packet length \\
$L_e$& Enlarged packet length (switching) \\
$g$& State variable, number of competitors\\
$k$& State variable\\
$p_c$    & Probability of unavailability due to PUs\\
$\chi$&Control channel availability per timeslot\\
$\eta$&Packet capture probability\\
$\psi$&Data channel availability $(1-p_c)*\eta$\\
$s_{max}$& Maximum No. of links equal to $\min(N, M_C)$\\
$H_{s_{max}}$& Probability of $k$ being equal to $s_{max}$\\
$P_s(g)$& Probability of success in Aloha-type competition \\
$P_s(k,g)$& Probability of success in state$(k,g)$ (\ref{eq:ps(k,g)}) \\
$P^m_s(k,g)$& Probability of success in state$(k,g)$ for a marked node\\
$p$& Probability of channel access attempt in Aloha-type competition\\
$q$& Packet length\\
${Qs}_{max}$& Maximum Queue Size\\
$T_k^{(j)}$& Probability of $j$ transmission among $k$\\
$S_g^{(j)}$& Probability of $j$ successful reservation\\
$Y_{a}^{b,c}$& Defined in (\ref{eq:Y_a^bc})\\
$\pi_{(k,g)}$& Steady-state probability of state $(k,g)$\\
$\pi^{N}_{(k,g)}$& Updated steady-state probabilities, eliminating states with $g=0$\\
$A_u$& Inter-arrival time of PUs (switching model)\\
\hline
\end{tabular}
\normalsize
\end{table}
\normalsize

\section{System Model}
\label {sec:system-model}
In this section we provide a summary of the two generic multichannel OSA MAC protocols for which we derive the delay analysis under unsaturated traffic in later sections. More details on those protocols, as well as their throughput performance under saturated traffic, can be found in \cite{pawelczak09,park11}.
It is assumed that there are $N$ nodes in an ad-hoc cognitive radio network and $M$ OSA channels where one channel is the dedicated control channel and the remaining $M_C=M-1$ channels are used for data transmission. Without loss of generality and for notation simplicity, we assume that the $N$ nodes have intended receivers that are not part of those nodes. 
A maximum of $s_{max}=\min(N,M_c)$ links might thus simultaneously exist. 

\begin{figure*}%
\centering 
\includegraphics[scale=0.62]{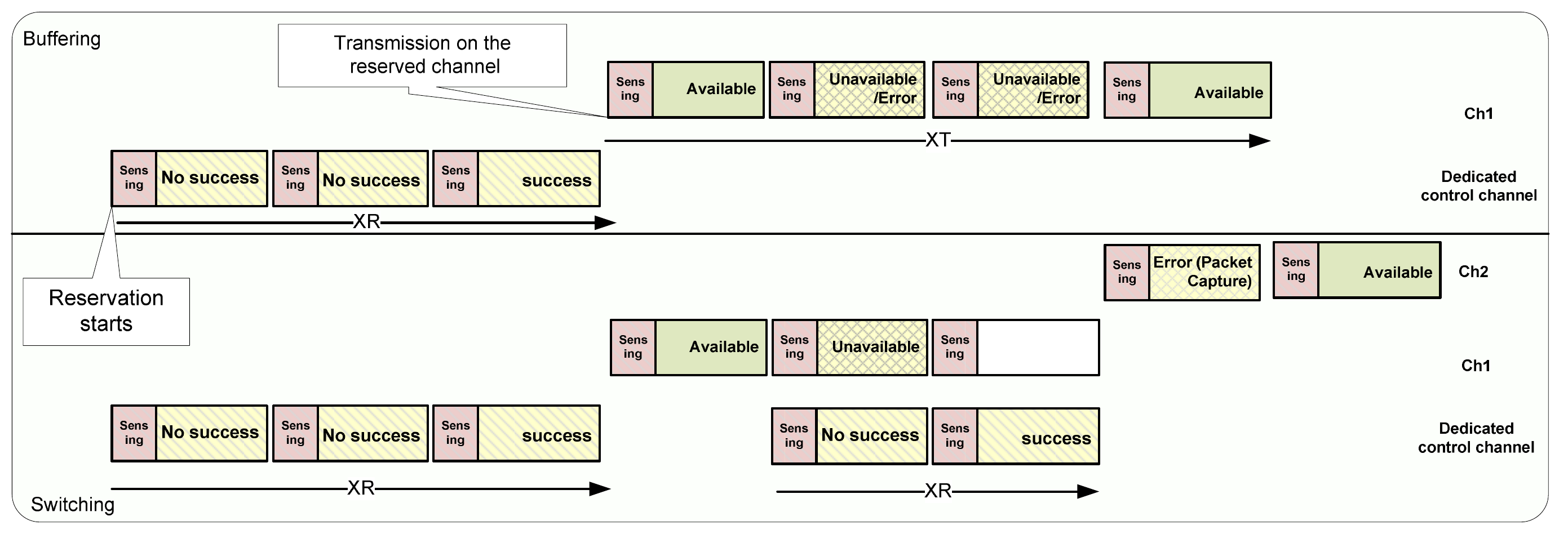}%
\caption{An example of the reservation and transmission processes. In the buffering model, a node reserves channel $Ch1$ in three timeslots ($X_R=3$). The transmission of a packet with the length 2 (timeslots) takes 4 timeslots to finish ($X_T=4$). The service time for this packet is thus $X=7$ timeslots. In the buffering model, after a channel unavailability in the second timeslot of the transmission on Ch1, the node participates in a new competition and reserve Ch2 after two timeslots. We thus have $X=8$.}%
\label{XTR.pdf}%
\end{figure*}

As illustrated in Fig.~\ref{XTR.pdf}, operations in the OSA network is time-slotted. Primary and secondary users are fully synchronized. 
In the beginning of each timeslot, there is a short quiet period during which the whole cognitive network stops operation to sense the channels. 
Primary users (PU) activity in any channel, including the control channel, is modeled by a Bernoulli random variable: in each timeslot 
and independent of the other timeslots and channels, a channel is unavailable for CR users (occupied by primary users or due to false alarms) 
with a probability $p_c$. We also use a packet capture model whereby a transmission during an available timeslot is successful, 
without considering collisions and interference, with  probability $\eta$ for the data channels and probability $\eta_C$ 
for the control channel. Channel availability and packet captures are assumed to be independent.

A geometric (Bernoulli) arrival process is assumed for all nodes with a probability of packet arrival in a timeslot $\lambda$. The packet length $L$ is given in the number of timeslots required to transmit the packet and has a geometric distribution with parameter $q$.

A node is in the busy state when it has a reserved data channel for transmission, otherwise the node is in the idle state (with an empty or non-empty packet queue). After the sensing period, if the control channel is available, the $g$ idle nodes with a non-empty packet queue attempt to reserve a channel using an Aloha-type competition over the control channel. That is, each of the $g$ nodes will transmit in the timeslot a reservation request with a probability $p$ over the control channel. The competition is successful if only one reservation request is transmitted and it is correctly received.
The probability of success in competition is given by~\cite{pawelczak09}:
\begin{equation}
P_s(g) =  g p (1-p)^{g-1} \chi,
\label{eq:Prob-Success}
\end{equation}
where $\chi=(1-p_c)\eta_C$ is the probability of a successful transmission on the control channel without considering collisions. The probability of success for a marked node among $g$ nodes is $\frac{P_s(g)}{g}$. 
It is assumed that the list of channels not being used by  CR nodes is given. So, a success in competition in this timeslot is followed by reserving the channel which is on the top of the list by the competition winner, to be used from the beginning of the next timeslot. An exception is when all channels are occupied in this timeslot. If a channel is released at the end of this timeslot, it will be assigned to the successful node in competition, otherwise the successful node in competition is unsuccessful in reserving a channel and returns for a new competition in the next timeslot. It is therefore important to distinguish the event of \emph{success in competition}, which depends solely on the number of competing users, from the event of \emph{success in reservation}, which depends also on the status of channels.
In case that a node is successful in reserving a channel, it starts transmitting one packet in the reserved channel from the beginning of the next timeslot. Other nodes can not make any interference on the reserved channel, but fading, sensing errors and PU activities may interrupt the transmission.

Two MAC protocols are considered in this paper. A  packet transmission cycle example for each protocol is illustrated in Fig.~\ref{XTR.pdf}. In the buffering MAC protocol~\cite{pawelczak09}, 
a node stays on its reserved channel, even when it becomes occupied by PUs, until the packet is entirely transmitted. 
In each timeslot, a successful transmission therefore occurs with probability $\psi=(1-p_c)\eta$. 
For the buffering MAC protocol, the (enlarged) packet service time $X$ therefore consists of a reservation period of length $X_R$ 
followed by a transmission time $X_T$ which consists of successful transmission, 
unsuccessful transmission and unavailable timeslots, until the entire packet is transmitted.

In the switching MAC protocol~\cite{park11}, a node that senses it channel occupied by PUs leaves the channel and returns to the idle state to participate in the competition to reserve a new channel in this timeslot. For the switching MAC protocol, the (enlarged) packet service time $X$ therefore consists of several reservation periods and the successful transmission and unsuccessful transmission timeslots required to transmit the entire packet.

 For both MAC protocols, once the packet transmission is terminated, the node releases the channel and returns to the idle state. It will enter the competition at the next timeslot if it has another packet in its queue. A service-resume transmission model is assumed for both MAC protocols where the remaining part of the packet is transmitted after each interruption. Note that due to the packet length geometric distribution model, the delay analysis is also the same for a service-repeat model where the entire packet must be transmitted after each interruption.

\section{Buffering MAC Protocol: Queue Occupancy Markov Chain Analysis}
\label{sec:Queue-Occupancy}
\begin{figure}%
\includegraphics[width=\columnwidth]{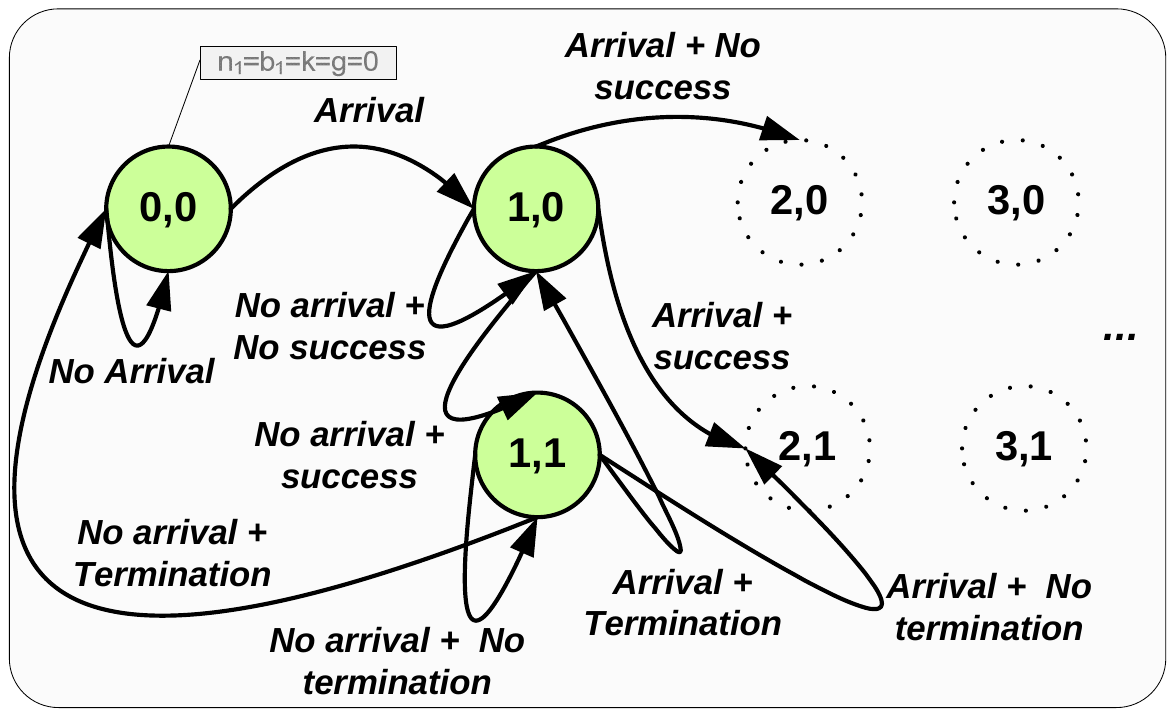}%
\caption{An example of the proposed queue occupancy Markov chain for $N=1$ and $M_C>0$.}%
\label{fig-Example-3-N1.pdf}%
\end{figure}

The discrete and memoryless nature of the system model and MAC protocols described in Section~\ref{sec:system-model} 
enables us to propose a general discrete-time Markov chain (MC) which tracks all arrivals, departures and 
interruption events for all nodes. The different queue performance metrics can then be directly computed using the 
steady-state MC solution.  The state variable is given in the form $(\vec{n},\vec{b})=(n_1,n_2,\dots,n_N,b_1,b_2,\dots,b_N)$ 
where $n_i \mbox { } (n_i=0,1,\dots)$ is the current number of packets in the system (waiting in the queue or being transmitted) 
of node $i$  and $b_i \mbox{ } (b_i=0,1)$ is a status flag indicating whether node $i$ has a reserved channel ($b_i=1$) or is 
idle ($b_i=0$). For each state $A$, we also define the variables $k_A=\sum{I_{(b_i=1, n_i>0)}}$ and $g_A=\sum{I_{(b_i=0, n_i>0)}}$, 
where $I$ is an indication function with $I_{true}=1$, $I_{false}=0$, which are, respectively, the total number of busy 
nodes and the total number of idle nodes with a non-empty queue. An MC example with $N=1$ is illustrated in Fig.~\ref{fig-Example-3-N1.pdf}. 

Because of the geometric arrival process and since at maximum one packet can be served in one timeslot 
over each channel, the queue status changes at most by one packet in two consecutive slots. 
That is $n_i(t)-1 \leq n_i(t+1) \leq n_i(t)+1$, where $n_i(t)$ is the node $i$ queue occupancy 
at time $t$. However, as a function of the status flag, the following refinements can be made:
\begin{enumerate}
	 \item If $n_i=0$ then $b_i=0$, and only an arrival event can occur, in which case $n_i(t+1)=1$, 
otherwise $n_i(t+1)=0$. Furthermore, the next status state is also $b_i=0$, irrespectively of an arrival event occurrence.
	\item If $n_i>0$ and $b_i=0$, two events of reservation success and an arrival can 
occur at the same timeslot and the events are independent. We have $n_i(t) \leq n_i(t+1) \leq n_i(t)+1$ 
depending on arrival or no arrival. Reservation success only changes the value of $b_i$ to 1.
	\item If $n_i>0$ and $b_i=1$, two events of packet transmission termination and an arrival can occur at the same timeslot and the events are independent. We have $n_i(t)-1 \leq n_i(t+1) \leq n_i(t)+1$. 
The case $n_i(t+1)=n_i(t)+1$ occurs with an arrival and no transmission termination, and the case $n_i(t+1)=n_i(t)-1$ 
occurs with no arrival and transmission termination. Meanwhile, a node will have the same number of packets in 
the queue in the next timeslot for two events: transmission termination and no arrival, or no transmission 
termination and no arrival. A transmission termination results in changing the value of $b_i$ to 0, irrespectively of 
an arrival event occurrence.
\end{enumerate}	
Furthermore, all busy nodes and idle nodes with an empty queue operate independently, 
so the joint probability of different events can easily be found by multiplication.

For two states $A=(\vec{n},\vec{b})=(n_1,\dots,n_N,b_1,\dots,b_N)$ and $B=(\vec{m},\vec{c})=(m_1,\dots,m_N,c_1,\dots,c_N)$,
$\mathbb{IS}$,  the set of impossible transitions, can then be defined as follows: 

\small
\begin{align}
&\mathbb{IS}=\Bigg{\{}(A,B)=((\vec{n},\vec{b}),(\vec{m},\vec{c}))|  \nonumber \\
&\exists n_i,m_i \mbox{ } (n_i=0 \mbox{ } \& \mbox{ } b_i=1) \mbox{ } \| (n_i=0 \& c_i=1) \mbox{ } \| \nonumber \\
&(m_i=0 \mbox{ } \& \mbox{ } c_i=1) \mbox{ } \| (Num_{s,AB}>1) \mbox{ } \| (m_i\notin  {(n_i-1){:}(n_i+1)}) \| \nonumber \\
&\bigg{(}\{k_A=s_{max}\} \& \{Num_{(s,AB)}>0\} \& \{Num_{(t,AB)}=0\}\bigg{)} \Bigg{\}}, 
\label{eq:}
\end{align}
\normalsize
where $Num_{(s,AB)}=|find(\vec{c}-\vec{b}==1)|$ ($|find(\vec{x}-\vec{y}==1)|$ 
is the number of positions for which the difference between the elements of $\vec{x}$ and $\vec{y}$ is 1)
is the number of nodes which 
started transmitting in the new timeslot from state $A$ to $B$ and $Num_{(t,AB)}=|find(\vec{b}-\vec{c}==1)|$ 
is the number of busy nodes in state A whose transmission is finished in state $B$.

The transition probabilities of going from state $A$ to a new state $B$ can then be computed using Alg.~({\ref{alg-queue-occ}}). For $(A,B) \in \mathbb{IS}$, the transition probability is zero. 
In Alg.~(\ref{alg-queue-occ}), $\mathbb{S}_g$ is the set of nodes participating in competition in 
state $A$ whose size is given by $|\mathbb{S}_g|$. If $\mathbb{S}_g$ is empty, the last part of the algorithm is not run. $AllBusyNoTer=1$ represents the case where all channels are busy and no termination occurs. 
\begin{algorithm}[H]
\caption{\scriptsize{The algorithm to find the transition probabilities $P_{A=(\vec{n},\vec{b})->B=(\vec{m},\vec{c})}$.}}
\label{alg-queue-occ}
\scriptsize
\begin{algorithmic}
\IF {$(k_A==s_{max}$ \& $Num_{(t,AB)}==0)$}
\STATE $AllBusyNoTer=1$
\ENDIF
\FOR{$i=1:N$}
 \IF {$(n_i==0)$}
    \STATE $P_i=\lambda^{m_i-n_i} (1-\lambda)^{1-(m_i-n_i)}$
\ELSE
    \IF {$b_i==1$}
					 \IF {$(c_i==1)$}
    					\STATE $P_i=(1-q\psi)\lambda^{m_i-n_i} (1-\lambda)^{1-(m_i-n_i)}$
					 \ELSE
    					\STATE $P_i=(q\psi)\lambda^{1-(n_i-m_i)} (1-\lambda)^{n_i-m_i}$
    			 \ENDIF
    \ELSE
    			 \STATE Put $i$ in $\mathbb{S}_g$
    			 \STATE $P_i=\lambda^{m_i-n_i} (1-\lambda)^{1-(m_i-n_i)}$
    \ENDIF
\ENDIF
 \ENDFOR
\STATE ***** Handling Competitions ***** $(\mathbb{S}_g \neq \emptyset)$
\STATE Find $j$ s.t.  $(c_{\mathbb{S}_g[j]}==1)$
	  \IF {$j>0$}
    	  \STATE $P_{\mathbb{S}_g[j]}=P_{\mathbb{S}_g[j]}* \frac{P_s(|\mathbb{S}_g|)}{|\mathbb{S}_g|}$
    \ELSE
				\IF {$(AllBusyNoTer==0)$}
							\STATE $P_{\mathbb{S}_g[1]}=P_{\mathbb{S}_g[1]}* [1-P_s(|\mathbb{S}_g|)]$
				\ENDIF
				
    \ENDIF


 \STATE FindProb($A$,$B$)=$\prod_{i}^{N}{P_i}$
\end{algorithmic}
\end{algorithm}
\normalsize

This exact Markov chain model is general and can address several system model extensions, 
such as non-homogeneous nodes with different arrival rates or different packet length. 
However, as there is no buffer limit, the number of states is infinite and grows exponentially 
with the number of nodes. To be able to solve the Markov chain numerically a buffer limit ${Qs}_{max}$ should be set
and the truncated Markov chain is analyzed instead. 
If the transition probabilities are written for a truncated version of the Markov chain with 
the buffer size ${Qs}_{max}$, the states with $\exists i \mbox{ s.t. } n_i={Qs}_{max}$ should 
be treated separately because for cases $b_i=0$, or $b_i=1$ and no termination, the queue 
occupancy in the next state with arrival or without any arrival (with probability equal to one) will still be $m_i={Qs}_{max}$.

Even for the truncated MC, the queue occupancy MC model suffers from exponential growth of the number states 
and is  thus not scalable. For example, for a buffer limit of ${Qs}_{max}$, the number of 
states will be $[2(1+{Qs}_{max})]^N$ (e.g., 484 states for $N=2$ and ${Qs}_{max}=10$). 
In the next section, the homogeneity of the nodes is taken into account to propose an approximate but scalable combined Markov chain which will be used for a service cycle analysis.

\section{Buffering MAC Protocol: Service Cycle Analysis}
\label{sec:Service-Cycle}
 
Since a geometric arrival process is assumed, we have for any 
node a discrete-time M/G/1 queue with geometric arrivals called Geom/G/1. Continuous-time M/G/1 and Geom/G/1 become equivalent when the timeslot is short with 
respect to the arrival rate. 
Furthermore, since 
the arrival process and packet length distribution are identical for all nodes, 
the queues are homogeneous. To solve a tagged node queue, we should therefore only find 
the service time $X=X_R+X_T$ for the buffering MAC protocol. $X_R$ and $X_T$ are independent, 
so the first and second moments of the service time $X$ can easily be found from the moments of $X_R$ and $X_T$. 
Discrete version of the {Pollaczek–-Khinchine} formula \cite{takagi93-2} can then be used to find the queue performance metrics of interest.
The exact distribution of $X_T$ and approximations for the distribution of $X_R$ are derived in Section~\ref{sec:tx} and~\ref{sec:xr}, respectively.

\subsection{Transmission Time}
\label{sec:tx}
For the buffering MAC protocol, the transmission time is started after the timeslot in which a channel has been reserved with success and lasts
until  the end of the packet transmission (see Fig. \ref{XTR.pdf}).
Since the node is using a reserved channel, there is no interference from other CR nodes and in each timeslot the probability of
a successful transmission is given by $\psi=(1-p_c)\eta$ (see Section~\ref{sec:system-model}).
For a packet with length $L=n$ timeslots, $n$ successful transmissions are required to complete
the packet transmission. For $X_T=k$ timeslots, the last slot should necessarily be a successful transmission and
among the remaining $k-1$ timeslots, exactly $n-1$ of them should be successful transmissions. Therefore, conditioned on the packet length $L=n$,
the transmission time of the packet  $X_T$  has a negative binomial distribution 
given by:
\begin{align}
&Pr(X_T=k|L=n)= \nonumber \\
&{k-1\choose n-1} \psi ^{n} (1-\psi)^{k-n}, \mbox{  } k=n,n+1,\dots
\label{eq:}
\end{align}
The packet length $L$ has a geometric distribution such that 
the unconditional distribution of $X_T$ is given by:
\begin{align}
&Pr(X_T=k) = \sum_{n} {Pr(X_T=k|L=n) f_L(n)} = \nonumber \\
&\sum_{n=1}^{\infty} {{k-1\choose n-1} \psi ^{n} (1-\psi)^{k-n} (1-q)^{n-1} q}.
\label{eq:}
\end{align}
The first and second moment of $X_T$ can then be found as:
\begin{equation}
E[X_T] = \frac{1}{q \psi} , \mbox{ } E[X^2_T] = \frac{2-q\psi}{(q \psi)^2}.
\label{eq:}
\end{equation}

\subsection{Reservation Periods}
\label{sec:xr}
At the beginning of every timeslot, nodes can be divided into three groups: 
busy nodes with a reserved channel, idle nodes with an empty queue 
and idle nodes with a non-empty queue. The probability of reservation success at the end of the timeslot 
depends on the number of nodes in the last group, except for the case where all channels are occupied. 
Due to packet arrival and packet transmission termination events, the number of idle nodes with a non-empty
queue changes from one timeslot to the next one such that the probability of 
reservation success in a timeslot is time-varying.
We will proceed as follows to find the reservation period distribution. First, we will propose a combined Markov chain to find 
the distribution and transition probabilities 
of the number of nodes competing in the reservation procedure. We will then find the reservation period distribution by first conditioning it 
on the number of competing nodes in the first reservation timeslot for the tagged node, and then using the steady-state probabilities to
obtain the unconditioned distribution.

\subsubsection{Combined Markov Chain}
\label{subsub:combined}
Since the sum of nodes at every timeslot is fixed to $N$, the state variable of the proposed combined Markov chain is given by a 2-tuple $(k,g)$ 
where $k$ is the number of busy nodes and $g$ is the number of idle nodes with a non-empty queue. 
The number of idle nodes with an empty queue is then given by $N-k-g$. 
We have the following facts about the system:
\begin{itemize}
	\item Every idle node with an empty queue may independently become a node with a non-empty queue in the next timeslot 
with the probability $P_a=\lambda$, which is the probability of one arrival during one timeslot. 
	\item Among the idle nodes with a non-empty queue, at most one node may become busy with the probability of reservation success $P_s(k,g)$ and
all others stay in the same group. $P_s(k,g)$ is the probability of getting a channel when $g$ nodes participate in competition and $k$ channels are already reserved. It is given by:
\begin{equation}
P_s(k,g)=\begin{cases}
P_s(g) & k<s_{max} \\
P_s(g) (1-T_k^{(0)})  & k=s_{max}\\
\end{cases}
\label{eq:ps(k,g)}
\end{equation}
where $T_k^{(0)}$ is the probability of no transmission termination among the $k$ ongoing communication sessions.
\item Every busy node may independently finish in a timeslot its packet transmission with  probability $q\psi$ and then, with probability $P_0$, the node 
joins the idle nodes with an empty queue. $1-P_0$ converges to the steady-state queue occupancy of a node queue and 
is the same for all busy nodes because nodes are homogeneous and have a geometric memoryless packet length. 
\end{itemize}
The number of participants in the competition in the next slot compared to the previous slot may thus decrease at most by one, 
but may increase to any larger value up to $N$ depending on the number of idle nodes with an empty queue with an arrival and the number of busy nodes that finish
their transmission and have a non-empty queue.
	 	
Using these facts and the results in \cite[Eq. (16)]{pawelczak09}, the transition probabilities of the combined Markov chain can be given by:

\scriptsize
\begin{align}
&P_{(k,g)->(z,h)} = \nonumber \\
&\begin{cases}
0& \mbox{ if } z>k+1, \forall g,h, \\
0& \mbox{ if } \forall k,z,  h < g-1, \\
T_k^{(0)}S_{g}^{(1)} Y_{(h-g+1)}^{(k-z+1),(N-k-g)}& \mbox{ if } z=k+1,  h \geq g-1,  \\
T_k^{(k-z)}S_{g}^{(0)} Y_{(h-g)}^{(k-z),(N-k-g)} + \\
T_k^{(k-z+1)}S_{g}^{(1)} Y_{(h-g+1)}^{(k-z+1),(N-k-g)}&0< z\leq k, k+z\neq 2s_{max}, h\geq g-1,\\
T_k^{(k-z+1)}S_{g}^{(1)} Y_{(h-g+1)}^{(k-z+1),(N-k-g)} +\\
T_k^{(k-z)}Y_{(h-g)}^{(k-z),(N-k-g)}& \mbox{ if } z=k=s_{max}, h\geq g-1,\\
T_k^{(k)}S_{g}^{(0)} Y_{(h-g)}^{(k-z),(N-k-g)}& \mbox{ if } z=0, h\geq g-1.\\
\end{cases}
\label{eq:comb_trans_prob}
\end{align}
\normalsize
$T_k^{(j)}$ and $S_g^{(j)}$ are respectively defined in \cite [Eqs. (14),(15)]{pawelczak09} 
as the probability of $j$ transmission terminations among $k$ ongoing communication sessions 
and the probability of $j$ success in competition among $g$ competitors (note that $S_g^{(1)}=P_s(g)$). 

$Y^{b,c}_{(a)}$ is the probability that $a$ new nodes join the group of idle nodes with a non-empty queue given that there are $b$ busy nodes 
that terminate their transmission and $c$ idle nodes with an empty queue. 
This happens if $i$ nodes out of the $b$ busy nodes 
that terminate their transmission have a non-empty queue at the end of the timeslot and $j$ nodes out of the
$c$ idle nodes with an empty queue have a packet arrival in the timeslot, where $i+j=a$. $Y^{b,c}_{(a)}$ is thus given by:

\footnotesize
\begin{align}
& Y_{a}^{b,c}= \nonumber \\
&\sum_{(i,j)|i+j=a, \mbox{ } 0\leq i\leq b \mbox{ $\&$ } 0\leq j\leq c} {{b\choose i} (1-P_0)^{i} (P_0)^{b-i} {c\choose j} (P_a)^{j} (1-P_a)^{c-j}}.
\label{eq:Y_a^bc}
\end{align}
\normalsize
Assuming that $P_0$ is known, the steady-state probabilities $\pi_{k,g}$  $\forall (k,g)$ of the combined Markov chain can be found. 
We explain in Section~\ref{sec:sol_proc} how $P_0$ can be found.

\subsubsection{Reservation Period Distribution}
\label{sec:res_per_dist}

We now derive the distribution of $X_R$ for a tagged node belonging to the group of idle nodes with a non-empty queue.
Let $X_R^{k,g}$ denote the reservation length given that the system is in the state $(k,g), g>0$ at the beginning of the reservation period.
The probability of a successful reservation for the marked node in the first timeslot, and thus have $X_R^{k,g}=1$ is equal to $P^m_s(k,g)=\frac{P_s(k,g)}{g}$. 
However, if the reservation is unsuccessful, the system transits to a new state $(z,h)$ with the probability $P_{(k,g)->(z,h)}|{(\text{No-Success})}$. 
In this state, the probability of successful reservation and thus have  $X_R^{k,g}=2$ is now $P^m_s(z,h)$. 
Otherwise, the system transits to a new state and the process repeats. In general, 
$\text{Pr}(X_R^{k,g}=i)$ can be found in a recursive manner from the probability of transition to other states $(z,h)$ and $\text{Pr}(X_R^{z,h}=i-1)$ from other states, given that the node was not successful in reservation. 

It is important to note that for this analysis, the transition probabilities of the combined MC discussed in the previous
section can not be directly used because it is implicitly assumed that the marked node was not 
successful. We therefore need a different MC with transition probabilities denoted by $Q$ conditioned on the fact
that at each transition, the marked node was not successful, and that if the reservation
competition was a success, it was one of the other $g-1$ competitors who was successful.  The transition probabilities $Q$
are given by (\ref{eq:comb_trans_prob}) where $P_s(g)$ given in (\ref{eq:Prob-Success}) is replaced by 
$(g-1) p (1-p)^{g-1} \psi$.

The distribution of $X_R^{k,g}$ is then given by:

\small
\begin{align}
&\text{Pr} (X_R^{k,g}=i)=  \nonumber \\
&\begin{cases}
P^{m}_s(k,g) & i=1 \\
(1-P^{m}_s(k,g))\sum_{(z,h)}{Q_{(k,g)->(z,h)}P^{m}_s(z,h)} & i=2\\
(1-P^{m}_s(k,g))\sum_{(z,h)}{Q_{(k,g)->(z,h)} (1-P^{m}_s(z,h))} & \\
\qquad \qquad \sum_{(a,b)}{Q_{(k,g)->(a,b)}P^{m}_s(a,b)} & i=3\\
....
\end{cases}
\label{eq:XRk,g}
\end{align}
\normalsize
$X_R^{k,g}$ can then be unconditioned to obtain $X_R$ as follows:
\begin{equation}
X_R=\sum_{(k,g),g>0}{X_R^{k,g} \pi^{N}_{(k,g)}}.
\label{eq:dist_xr}
\end{equation}
Since when we mark a node to find $X_R$, 
we implicitly assume that the node is idle with a non-empty queue,
the states in which the number of competing users is zero ($g=0$) 
should therefore be eliminated to obtain the steady-states probabilities for the states $(k,g), g>0$ as $\pi^{N}_{(k,g)}= 
\frac{\pi_{(k,g)}}{1-\sum_{(k,g)|g=0}{\pi_{(k,g)}}}$.

\subsubsection{Solution Procedure}
\label{sec:sol_proc}

To find the combined MC transition and steady-state probabilities, $P_0$, the steady-state probability of 
a node's queue being empty, is required. $P_0$ can be computed from the Geom/G/1 queue
relation $(1-P_0)=\lambda E[X]=\lambda E[X_R+X_T]$. However, $E[X_R]$ depends on the combined MC solution.  
We therefore proposed an iterative solution where an initial value is set for $P_0$. 
The combined MC chain can then be solved and $E[X_R]$ computed. A new value of $P_0$ is then obtained and
used for the next iteration.
The iteration continues until the value of $P_0$ converges. 
To compute the initial value of $P_0$, we propose to use the lower bound case for $X_R$  
where only the tagged node participates in the competition. The distribution of $X_R$ is then geometric and $E[X_{R}] = (p \chi)^{-1}$.

Note that the number of competing nodes in one state is not independent of the other states. 
However by using the steady-state probabilities in (\ref{eq:dist_xr}) combined with the iterative solution procedure, this dependency is ignored. 
Furthermore, the different functions involve in this system are not necessarily convex. 
The combined-MC approach is therefore inherently an approximate solution.
For example, we have observed that in some cases the combined-MC solution converges to a non-zero value of $P_0$ while
simulations indicated that the system is unstable. However, all studies that we have performed showed that in the region where
the system is stable, the combined-MC approximation is very accurate (see results presented in Sec.~\ref{sec:simulation}).

\subsubsection{Other Approximations}
\label{sec:xr_approx}

The proposed approach based on the combined Markov chain is scalable. 
However, the number of slots until 
reservation success is infinite and the distribution of $X_R$ should be truncated
which results in an approximation. Furthermore, even when the distribution of $X_R$ is truncated,
the computations are intensive. We therefore propose in this section three different approaches
to reduce the computational complexity required to calculate the moments of $X_R$.

\paragraph{Reduced Markov Chain}
The number of states in the Markov chain can be reduced by assuming
that idle nodes will independently have a non-empty queue at the beginning of
a timeslot, and therefore participate in the competition, with probability $1-P_0$.
This assumption therefore ignores the dependency between timeslots of the
status of an idle node and we don't need to track in the MC state variable
the number of idle nodes with a non-empty queue. That is, the combined MC chain presented
in Section~\ref{subsub:combined} can be replaced with the single dimension 
Markov chain proposed in \cite{pawelczak09} which uses the single state variable $k$, the number
of busy nodes. 
In each timeslot, $N-k$ nodes then participate in the competition with an
Aloha access probability given by $p_m=p (1-P_0)$. 
The distribution of $X_R$ can then be computed as explained in Section~\ref{sec:res_per_dist}  where
$P_s(g)$ given in (\ref{eq:Prob-Success}) is replaced by $(N-k-1) p_m (1-p_m)^{N-k-1} \psi$,
the summations in (\ref{eq:XRk,g}) and (\ref{eq:dist_xr}) are over a single variable, and the modified transition probabilities in (\ref{eq:XRk,g}) and
steady state probabilities in (\ref{eq:dist_xr}) are obtained from the MC in~\cite{pawelczak09}.  
The solution procedure is then the same as the one described in Section~\ref{sec:sol_proc}.
This approximate model, denoted by 'Pawelczak-MC' in the numerical results, reduces the computational complexity since the state space is smaller.

\paragraph{Approximate $X_R$ Distribution}
\label{Avg}
We now propose an approximate approach which significantly reduces the complexity associated with finding
the distribution of $X_R$ so that it can be used in scenarios with limited computation capacity. 
The  complexity in (\ref{eq:XRk,g}) arises from the fact that the reservation success probability is time-varying
and has memory. As an approximation, we propose to simply assume that the number of participating nodes during a reservation period is always
$n$.  With this assumption, $X_{nR}$, the reservation time given that the number of participating nodes is $n$ at each timeslot,
has a geometric distribution with probability 
$(1-H_{s_{max}})P_s^m(k,n|k<s_{max})+H_{s_{max}}P_s^m(k=s_{max},n)$,
where $H_{s_{max}}$ is the probability of $k$ being equal to $s_{max}$ and is defined as follows:
\begin{equation}
H_{s_{max}}= \sum_{(k,g)|k=s_{max}}{\pi^{N}_{(k,g)}}.
\label{eq:}
\end{equation}
From the combined MC chain presented in Sec.~\ref{subsub:combined}, we can compute
$Pr(g=n|g \neq 0)$ ($n=1,\dots,N$), the distribution of the number of nodes participating in the competition
excluding the cases where there is no competitors. We can then approximate the distribution of $X_R$ as follows:
\begin{equation}
X_R \approx \sum_{n} Pr(g=n|g \neq 0) X_{nR}.
\label{eq:}
\end{equation}
Given that $X_{nR}$ is geometric, it is then easy to compute the moments of $X_R$.
Since $P_0$ is not known, 
the same solution procedure as the one described in Section~\ref{sec:sol_proc} should be used.
This approximate model is denoted by Combined-MC-Dist in the numerical results.

\paragraph{Average $X_R$ Distribution}
This third approach further simplifies the computations by calculating the reservation time with the average number of idle nodes in competition with a non-empty queue.
Let define 
\begin{equation}
\bar{G} = \sum_{(k,g)|g\neq 0} {g \pi^{N}_{k,g}}.
\label{eq:}
\end{equation}
where the steady-state probabilities $\pi^{N}_{k,g}$ are obtained from the combined MC chain presented in Section~\ref{subsub:combined}.
Then $X_R$ is approximated with a geometric distribution with 
probability $(1-H_{s_{max}})P_s^m(k,\bar{G}|k<s_{max})+H_{s_{max}}P_s^m(s_{max},\bar{G})$.
Since $P_0$ is not known, 
the same solution procedure as the one described in Section~\ref{sec:sol_proc} should be used.
This approximate model is denoted by Combined-MC-Avg in the numerical results.

\section{Switching MAC Protocol Delay Analysis}
\label{sec:switching}

In this section, we provide the delay analysis for the multichannel OSA switching MAC protocol. 
The same approach as for the delay analysis for the buffering MAC protocol is followed: we first 
provide a system occupancy Markov chain in Section~\ref{sec:switching_occ_mc} and then we derive an Geom/G/1 queue analysis 
in Section~\ref{sec:switching_service_cycle}. 
The only difference between the buffering and switching MAC protocols is that in the switching protocol, if a SU senses that its reserved channel is occupied
by a PU, it immediately returns to the group of idle nodes with non-empty queue and participates in the competition, while in the buffering MAC protocol
the node would remain on its reserved channel. To facilitate the analysis of the switching MAC protocol, the system occupancy MC and combined MC for service time analysis
are embedded after the sensing period. Finally, for
sake of brevity, we omit details in the analysis that are similar to the buffering MAC protocol analysis.


\subsection{Occupancy Markov Chain}
\label{sec:switching_occ_mc}

The states of the queue occupancy Markov chain are similar 
to the buffering case and are given by $(\vec{n},\vec{b})$. 
However, the observation time where the Markov chain has been 
imbedded is immediately after the initial sensing period when 
the status of the channels has already been determined. 
We therefore have that when $b_i=1$ the reserved channel for node $i$
is available in this timeslot for transmission (the transmission may however be unsuccessful due to the packet capture model).

\begin{figure}%
\includegraphics[width=\columnwidth]{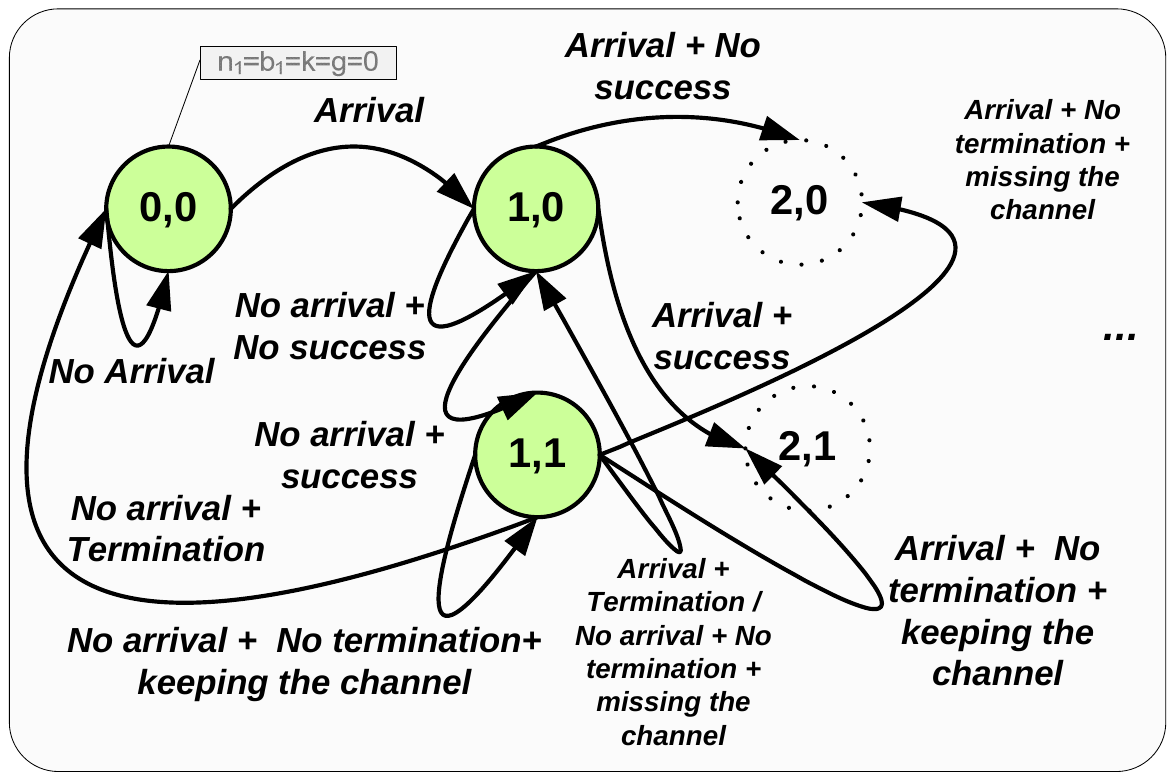}%
\caption{An example of the proposed queue occupancy Markov chain for switching policy when $N=1$ and $M_C>0$.}%
\label{fig-Example-3-N1-Switching.pdf}%
\end{figure}

Similar to the buffering case, for a state $A$, we can define $k_A$ as the 
total number of busy nodes who have a channel and whose channels 
are available in this timeslot and $g_A$ as the total number of idle nodes with non-empty queue. 
The switching model facts are very similar to the buffering model with the difference 
that a node who transmitted on its assigned channel
in this timeslot may, if it does not finish its packet transmission,
 come back to the competition with probability $p_c$ if a PU is detected on its assigned channel.
For $N=1$, the Markov chain is illustrated in Fig.~\ref{fig-Example-3-N1-Switching.pdf} 
where the transitions after states $(2,0)$ and $(2,1)$ are not shown. 
It can be seen that compared to Fig.~\ref{fig-Example-3-N1.pdf} for the buffering policy, 
there are only two changes: a new transition may occur from state (1,1) to (1,0) 
if no arrival occurs, transmission is not finished, but the channel is missed in the next timeslot, 
and to state (2,0) with the same conditions and an arrival occurs.

Transition probabilities for this Markov chain can be found 
using Alg. \ref{alg-queue-occ-switching} with the same set of impossible states (IS). 
Similar to the buffering model, when a node $i$ is busy ($b_i=1$), 
its transitions are independent of other nodes. For $b_i=0$ and $n_i>0$, 
the node participates in competition. 
Compared to the buffering model, it can be seen that 
the main change is for the case where there is a transition between $b_i=1$ and $c_i=0$ 
because this transition may occur in the switching model due to both a service termination 
or a PU positive sensing. Moreover, we can see that the probability 
of success is updated by $1-p_c$ because a success in competition 
should also be followed by the availability of the reserved channel to be a reservation success.

For the case where we have $k=s_{max}$, the success depends on
the nodes who had a transition from $b_i=1$ to $c_i=0$.
Any of them may have a transition due to missing the channel and PU arrival (kept in vector $P_{m}$) 
or a termination (kept in vector $P_{t}$). 
$P_{mt}$ keeps the total probability of both events. Then, $P_{Allm}$  
is defined to be the probability that all transitions from $b_i=1$ 
to $c_i=0$ have been due to PU arrival. In this special case, 
no reservation success is possible. If at least one node among 
them has finished its transmission, with the probability $P_{10T}$, 
then a competition success may result into a reservation success with the probability $1-p_c$.

\begin{algorithm}
\caption{\scriptsize{The algorithm to find the transition probabilities (FindProb($A$,$B$)) in the switching model. Note that $f^0_{ap}(n_i,m_i)=\lambda^{m_i-n_i} (1-\lambda)^{1-(m_i-n_i)} \mbox{ } (m_i \geq n_i)$ and $f^1_{ap}(n_i,m_i)=\lambda^{1-(n_i-m_i)} (1-\lambda)^{n_i-m_i} \mbox{ } (m_i \leq n_i)$.}
\label{alg-queue-occ-switching}}
\tiny
\begin{algorithmic}
\IF {$(k_A==s_{max}$ \& $Num_{(t,AB)}==0)$}
\STATE $AllBusyNoTer=1$
\ENDIF
\FOR{$i=1:N$}
 \IF {$(n_i==0)$}
    \STATE $P_i=f^0_{ap}(n_i,m_i)$
\ELSE
    \IF {$b_i==1$}
					 \IF {$(c_i==1)$}
    					\STATE $P_i=(1-q\eta)(1-p_c)f^0_{ap}(n_i,m_i)$
					 \ELSE
					 			\IF {$(k_A!=s_{max})$}
    								\STATE $P_i=(1-q\eta)(p_c)f^0_{ap}(n_i,m_i)+(q\eta)f^1_{ap}(n_i,m_i)$
    						\ELSE
             			  \STATE Put $i$ in $\mathbb{S}_{10}$
             			  \STATE $h=h+1$
    								\STATE $P_i=1$
    								\STATE $P_{t,h}=(q\eta)f^1_{ap}(n_i,m_i)$
    								\STATE $P_{m,h}=(1-q\eta)(p_c)f^0_{ap}(n_i,m_i)$
    								\STATE $P_{mt,h}=(1-q\eta)(p_c)f^0_{ap}(n_i,m_i)+(q\eta)f^1_{ap}(n_i,m_i)$
    						\ENDIF
					 					
    			 \ENDIF
    \ELSE
    			 \STATE Put $i$ in $\mathbb{S}_g$
    			 \STATE $P_i=f^0_{ap}(n_i,m_i)$
    \ENDIF
\ENDIF
 \ENDFOR
\FOR{$i=1:|\mathbb{S}_{10}|$}
 \STATE $P_{10T}=P_{10T}+ \prod_{1}^{h-1}{P_{m,h}} P_{t,h} \prod_{h+1}^{|\mathbb{S}_{10}|}{P_{mt,h}}$
\ENDFOR
 \STATE $P_{Allm}=\prod_{1}^{|\mathbb{S}_{10}|}{P_{m}}$
\STATE FindProb($A$,$B$)=$\prod_{i}^{N}{P_i} \prod{P_{mt,h}}$
\STATE ***** Handling Competitions ***** $(IF \mathbb{S}_g \neq \emptyset)$
\IF {$(k_A!=s_{max})$}
		\STATE Find $j$ s.t.  $(c_{\mathbb{S}_g[j]}==1)$
	  \IF {$j>0$}
    	  \STATE $P_{\mathbb{S}_g[j]}=P_{\mathbb{S}_g[j]}* \frac{P_s(|\mathbb{S}_g|)}{|\mathbb{S}_g|}(1-p_c)$
    \ELSE
				\IF {$(AllBusyNoTer==0)$}
							\STATE $P_{\mathbb{S}_g[1]}=P_{\mathbb{S}_g[1]}* [1-P_s(|\mathbb{S}_g|)(1-p_c)]$
				\ENDIF
				
    \ENDIF

\ELSE
		\STATE Find $j$ s.t.  $(c_{\mathbb{S}_g[j]}==1)$
	  \IF {$j>0$}
    	  \STATE $P_{\mathbb{S}_g[j]}=P_{\mathbb{S}_g[j]}* \frac{P_s(|\mathbb{S}_g|)}{|\mathbb{S}_g|}(1-p_c)$
    	  \STATE $P_{\mathbb{S}_m[1]}=P_{10T}$
    \ELSE
				\IF {$(AllBusyNoTer==0)$}
							\STATE $P_{\mathbb{S}_g[1]}=P_{\mathbb{S}_g[1]}* [P_{Allm}+P_{10T}(1-P_s(|\mathbb{S}_g|))+P_{10T}P_s(|\mathbb{S}_g|)(1-p_c)]$
				\ENDIF
				
    \ENDIF
\ENDIF

\STATE FindProb($A$,$B$)=$\prod_{i}^{N}{P_i}$
\end{algorithmic}
\end{algorithm}
\normalsize

\subsection{Service Cycle Analysis}
\label{sec:switching_service_cycle}

\begin{figure}%
\includegraphics[width=\columnwidth]{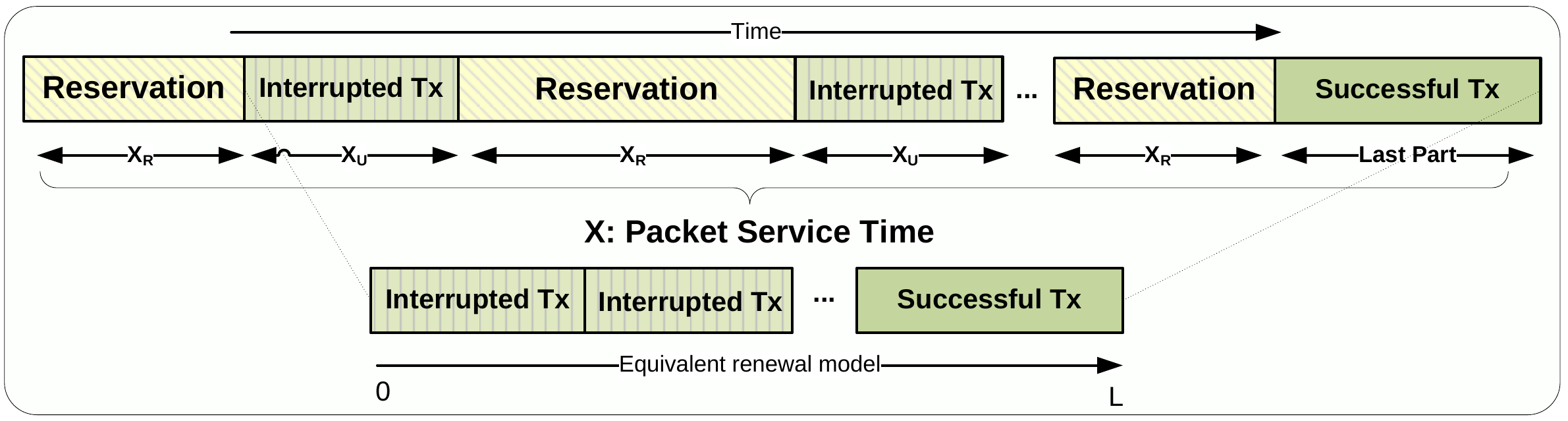}%
\caption{Unsuccessful transmission and reservation periods create alternating renewal processes.}%
\label{fig-B0S0-model.pdf}%
\end{figure}

The nodes queue in the multichannel OSA network with the switching MAC protocol can 
also be modeled with a Geom/G/1 queue. 
As illustrated in Fig.~\ref{fig-B0S0-model.pdf}, the Geom/G/1 service time in this queue is composed of alternating renewal processes: 
uncompleted packet transmissions (a part of the packet however can be transmitted) 
and competitions for channel reservation alternate until the packet is entirely transmitted. 
Since the packet length is geometric and the occurrence of an interruption is a Bernoulli event, and both are independent, the uncompleted packet transmission time is thus memoryless and identical. The unsuccessful transmission attempts therefore form a renewal process.
The reservation periods $X_R$ during the transmission 
of a packet are also identical and form the second renewal process. 
Let $X$ be the service time of the Geom/G/1 queueing model, which can be given by
\begin{equation}
X=X_R+L_e+n X_R = L_e+ m X_R,
\label{eq:}
\end{equation}
where $L_e$ is the enlarged packet length (in number of slots), 
which is geometrically distributed with 
the probability $q \eta$, 
and $n=m-1$ is the number of switching during the transmission 
of the packet. To find the service time, we must find the distribution of $X_R$ and the number of renewals $n$, 
which are addressed in Section~\ref{sec:switching_X_R} and Section~\ref{sec:switching_renewal}, respectively.

\subsubsection{Reservation Period}
\label{sec:switching_X_R}

The approach to find the distribution of $X_R$ for the switching MAC protocol is similar as the one followed
for the buffering model in Sec.~\ref{sec:xr}.
The state variable of the combined Markov chain is still 
given by a 2-tuple $(k,g)$, 
however,  the observation time where the combined Markov chain is 
embedded for the switching MAC protocol is after the 
initial sensing period at the beginning of each timeslot such that the status of the channels 
for the current timeslot is known. 
$k$ is therefore the number of reserved channels
on which there will be a transmission attempt in this timeslot and $g$ is the number
of idle nodes 
with a nonempty queue which participate in the competition on the control channel for reservation.
The number of idle nodes with an empty queue is thus given 
by $N-g-k$. 

To find transition probabilities, we use four 
auxiliary variables $i$, $j$, $n$ and $m$. Note that in 
the following,  $k$ and $g$ refer to $k(t)$ and $g(t)$ 
for the current timeslot $t$. 
Among $k$ busy nodes, $i$ nodes independently finish the transmission 
of their packets, each with the probability $q \eta$. 
We use the notation $T_{k}^{i}={k \choose i} q\eta^{i} (1-q\eta)^{k-i}$ 
for the probability of this event. Any of those $i$ nodes may  join in the next timeslot, with probability $1-P_0$, the idle nodes with a nonempty 
queue. Otherwise, they join 
the idle nodes with an empty queue. 
The probability that $n$ nodes out of the $i$ finishing nodes will be among the $g(t+1)$ nodes in the next
timeslot is thus given by  ${i \choose n} {P_0}^{i-n} (1-P_0)^{n}$.
For the remaining $k-i$ nodes which do not finish their transmission, $j$ nodes 
may leave their channels due to PU presence sensing with  
probability $F^{k-i-j}_{k-i}={k-i \choose j} {1-p_c}^{k-i-j} (p_c)^{j}$.
Those nodes also join the $g(t+1)$ idle nodes with a nonempty queue after the sensing period.
The remaining $k-i-j$ nodes stay on their channels, 
so they are among the $k(t+1)$ nodes.

Considering $g$ participants in competition, we have the 
same results for competition success as for the buffering case. One node may then be 
successful and join the $k(t+1)$ nodes with probability $P_s(k,g) (1-p_c)$. The 
$1-p_c$ term is included because the node may be successful in 
competition and reservation during this timeslot, but 
the reserved channel might be unavailable in the next timeslot. 
Therefore, without starting the transmission, the node 
has to come back to competition.  All other $g-1$ nodes 
remain in the group of idle nodes with a nonempty queue, 
i.e., they are among $g(t+1)$ nodes. Finally, any of 
$N-k-g$ idle nodes with an empty buffer may receive a 
packet and join the participants in competition in the 
next timelot. The probability to have $m$ such nodes is 
given by ${N-k-g \choose m} {\lambda}^{m} (1-\lambda)^{N-k-g-m}$. 

We can thus, for the switching MAC protocol, express the transition probabilities 
of the combined Markov chain in the form $P_{(k,g)->(z,h)}$ as follows:

\begin{figure*}[!t]
\setcounter{mytempeqncnt}{\value{equation}}
\setcounter{equation}{14}
\footnotesize
\begin{align}
&P_{(k,g)->(z,h)}= \nonumber \\
&\begin{cases}
0& \mbox{ if } z>k+1, \forall g,h, \\
0& \mbox{ if } \forall k,z,  h < g-1, \\
T_k^{(0)} F^{k}_{k} S_{g}^{(1)}{(1-p_c)} Y_{(h-g+1)}^{(k-z+1),(N-k-g)}& \mbox{ if } z=k+1, \\
\sum_{i=0}^{k-z}{T_k^{i} F^{z}_{k-i} (1-S_{g}^{(1)}(1-p_c)) Y_{(h-g-(k-i-z))}^{i,(N-k-g)}} + 
\sum_{i=0}^{k-z+1}{T_k^{i} F^{z-1}_{k-i} (S_{g}^{(1)}(1-p_c)) Y_{(h-g-(k-i-z))}^{i,(N-k-g)}} & \mbox{ if } 0<z\leq k, k+z\neq 2s_{max}, \\
\sum_{i=0}^{k-z}{T_k^{i} F^{z}_{k-i} Y_{(h-g-(k-i-z))}^{i,(N-k-g)}} + 
\sum_{i=0}^{k-z+1}{T_k^{i} F^{z-1}_{k-i} (S_{g}^{(1)}(1-p_c)) Y_{(h-g-(k-i-z))}^{i,(N-k-g)}} & \mbox{ if } z=k=s_{max},\\
\sum_{i=0}^{k-z}{T_k^{i} F^{z}_{k-i} (1-S_{g}^{(1)}(1-p_c)) Y_{(h-g-(k-i-z))}^{i,(N-k-g)}}& \mbox{ if } z=0,
\end{cases}
\label{eq:}
\end{align}
\normalsize
\setcounter{equation}{\value{mytempeqncnt}}
\hrulefill
\vspace*{4pt}
\end{figure*}
%
where $Y_{a}^{b,c}$ is given in (\ref{eq:Y_a^bc}).
Given the transition probabilities for the combined Markov chain for the switching MAC protocol, the distribution of $X_R$ can
be obtained as explained in Section~\ref{sec:res_per_dist}. The second and third approximations given
in Section~\ref{sec:xr_approx} to find the distribution of $X_R$ can also be used (the approximation based on the
model presented in~\cite{pawelczak09} is not applicable for the switching MAC protocol).

\subsubsection{Renewal Process}
\label{sec:switching_renewal}

As illustrated in Fig.~\ref{fig-B0S0-model.pdf},
$X_{U}$, any part of the whole service time where the node has a reserved
channel, is the inter-arrival time of the channel unavailability 
events.
Let us use the random variable 
$A_u$ to denote the inter-arrival time of those events. $A_u$ has 
a geometric distribution with probability $p_c$.
From the 
Renewal Theory point of view~\cite{cox62}, $n=m-1$ is the 
number of renewals from 0 to $L_e$ where the process which is 
renewed is $A_u$.
The number of renewals can be found from Renewal Theory 
results \cite{cox62}. 

However, as we have  geometric 
distributions for both $L_e$ and $A_u$, we can use the following simpler 
approach.
Due to the memoryless packet length, the remaining part of the packet in any operating period is geometrically 
distributed with parameter $q$.  The total number of renewals is 
thus the number of trials until success, which is a geometric distribution of type-II, with the 
probability of success in a period being $Pr(L_e<=A_u)$, the probability of sending the 
whole packet in the given period. However, the number 
of reservation periods which appear in the calculation of service 
time is $m=n+1$ because the first reservation period is not counted as a 
renewal (see Fig.~\ref{fig-B0S0-model.pdf}). We thus have:
\begin{equation}
Pr(n==k)=(1-Pr(L_e<=A_u))^{k} Pr(L_e<=A_u).
\label{eq:}
\end{equation}
For any two type-I geometrically distributed random 
variables $R_1$ and $R_2$ with parameters $p_1$ and $p_2$, we can show that
\begin{equation}
Pr(R_1<=R_2) = \frac{p_1}{1-(1-p_1)(1-p_2)}.
\label{eq:}
\end{equation}
We thus have:
\begin{equation}
Pr(L_e<=A_u)= \frac{q\eta}{1-(1-q\eta)(1-p_c)},
\label{eq:}
\end{equation}
\begin{equation}
E[n]=\frac{1-\frac{q\eta}{1-(1-q\eta)(1-p_c)}}{\frac{q\eta}{1-(1-q\eta)(1-p_c)}}=\frac{p_c(1-q\eta)}{q\eta},
\label{eq:}
\end{equation}
and
\begin{equation}
E[n^2]=(E[n])^2+\frac{1-\frac{q\eta}{1-(1-q\eta)(1-p_c)}}{[\frac{q\eta}{1-(1-q\eta)(1-p_c)}]^2}.
\label{eq:}
\end{equation}

To find the second moment of $X$ it should be taken into account that we have a 
random sum and the two parameters $n$ and $L_e$ are not independent. Using the 
following facts: 
\begin{equation}
\sum_{k=1}^{\infty} k(1-q)^{k-1} = \frac{1}{q^2},
\label{eq:}
\end{equation}
\begin{equation}
\sum_{k=1}^{\infty} k(k-1)(1-q)^{k-2} = \frac{2}{q^3},
\label{eq:}
\end{equation}
and thus
\begin{equation}
\sum_{k=1}^{\infty} k^2(1-q)^{k-2} = \frac{2-q}{q^3(1-q)},
\label{eq:}
\end{equation}
the term $E[L_en]$ can be given by:
\footnotesize
\begin{align}
&E[L_en]= \nonumber \\
&\sum_{k=1}^{\infty}{E[L_en|(L_e=k)] Pr(L_e==k)}=\sum_{k=1}^{\infty}{k E[n|(L_e=k)] Pr(L_e==k)} \nonumber \\
&=\sum_{k=1}^{\infty}{k (k-1) p_c q\eta(1-q\eta)^{k-1}} = \frac{2p_c(1-q\eta)}{(q\eta)^2}.
\label{eq:cor-of-T-Ka}
\end{align}
\normalsize
We have used the fact that the number of renewals in the 
interval $[0 \mbox{ } k]$ of a type-I geometric with the parameter $p_c$ is equal to $(k-1)p_c$.

From the moments of $n$ and $X_R$, it is then straightforward to find the moments of $X$ and solve the Geom/G/1 queue.
Note that since $P_0$ is initially unknown and is required to find the distribution of $X_R$, the solution
procedure outlined in Section~\ref{sec:sol_proc} can be followed. The initial value of $P_0$ is computed with the lower bound of $X_R$ obtained for the case where only a
single node is competing and for which $E[X_{R}] = (p \chi(1-p_c))^{-1}$.

\section{Simulation and Numerical Results}
\label{sec:simulation}

In this section, we verify the accuracy of the proposed 
models and approximations by comparing simulation results with numerical evaluations
of the theoretical results presented in this paper. We also study the impact of several parameters on the
multi-channel OSA MAC protocols performance.
Simulations are done with Matlab where each simulation scenario is 
run for 350000 timeslots and the same scenario is repeated 10 times.
Unless indicated otherwise, we have $M_C=N$ data channels. In Sec.~\ref{sec:res_arr_rate} to \ref{sec:res_tx_cap}, we investigate
the buffering MAC protocol while in Sec.~\ref{sec:sim-switch} we present results for the switching MAC protocol.

\subsection{Impact of Arrival Rate}
\label{sec:res_arr_rate}

\begin{figure*}%
\includegraphics[width=0.41\columnwidth]{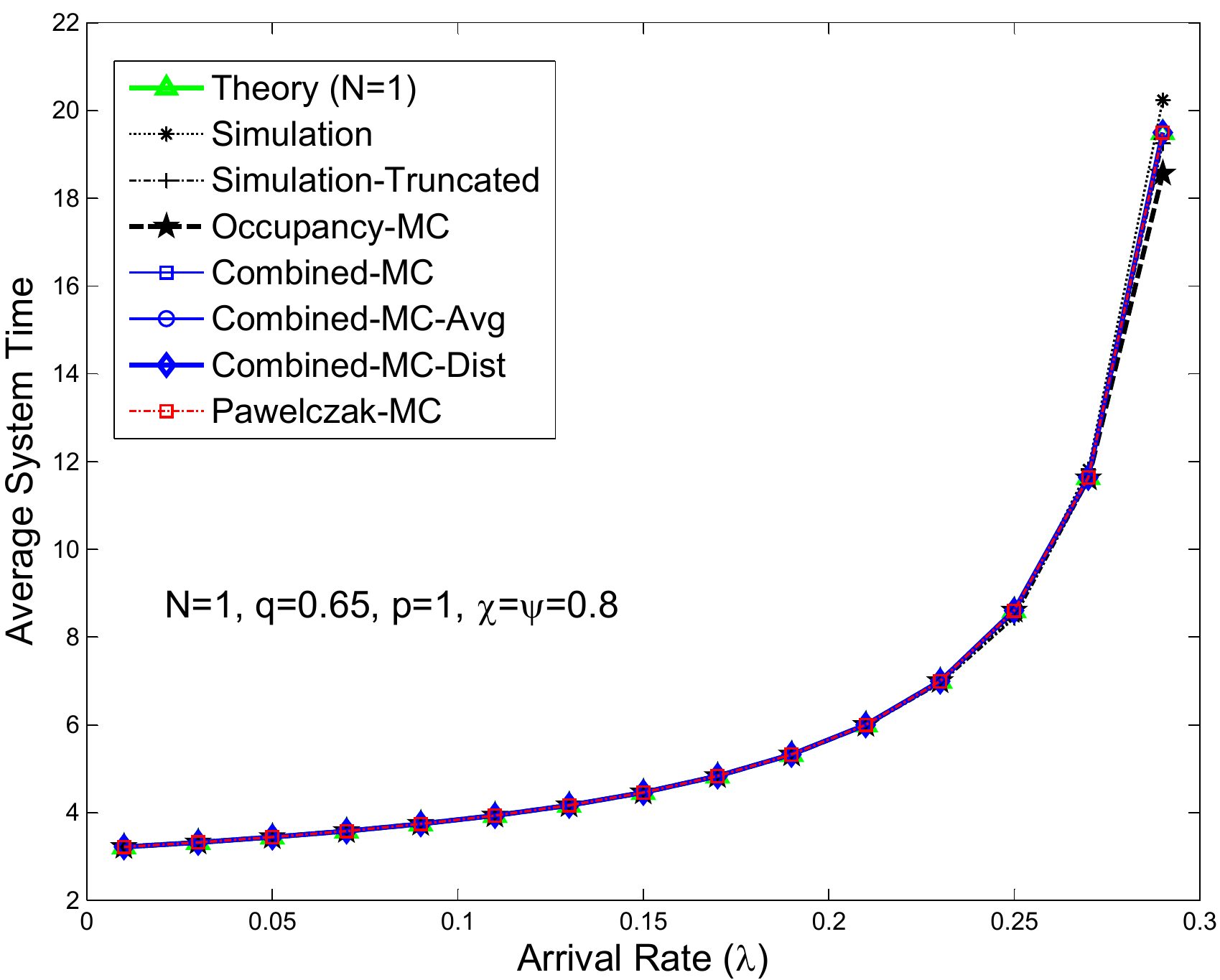}%
\includegraphics[width=0.41\columnwidth]{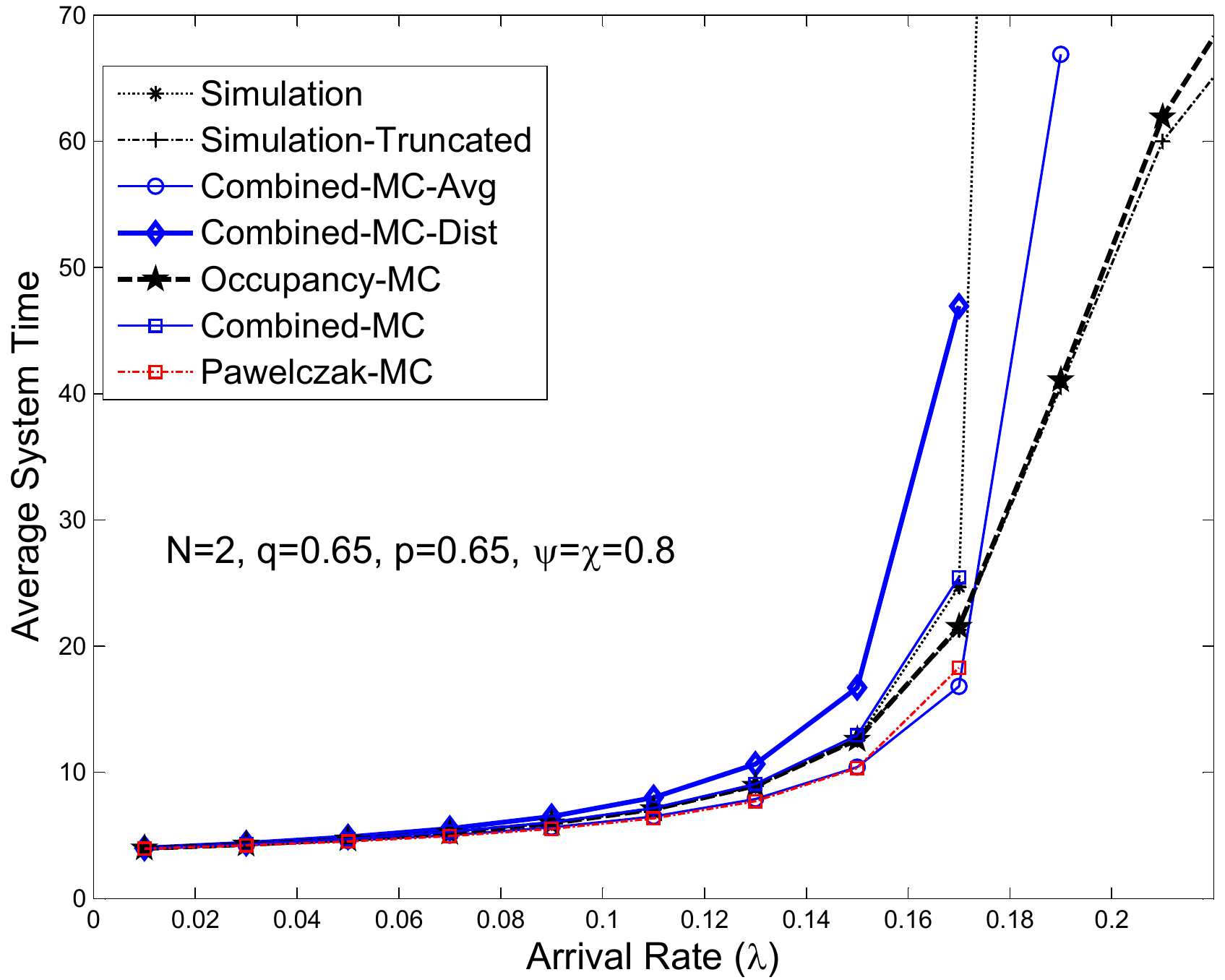} 
\includegraphics[width=0.41\columnwidth]{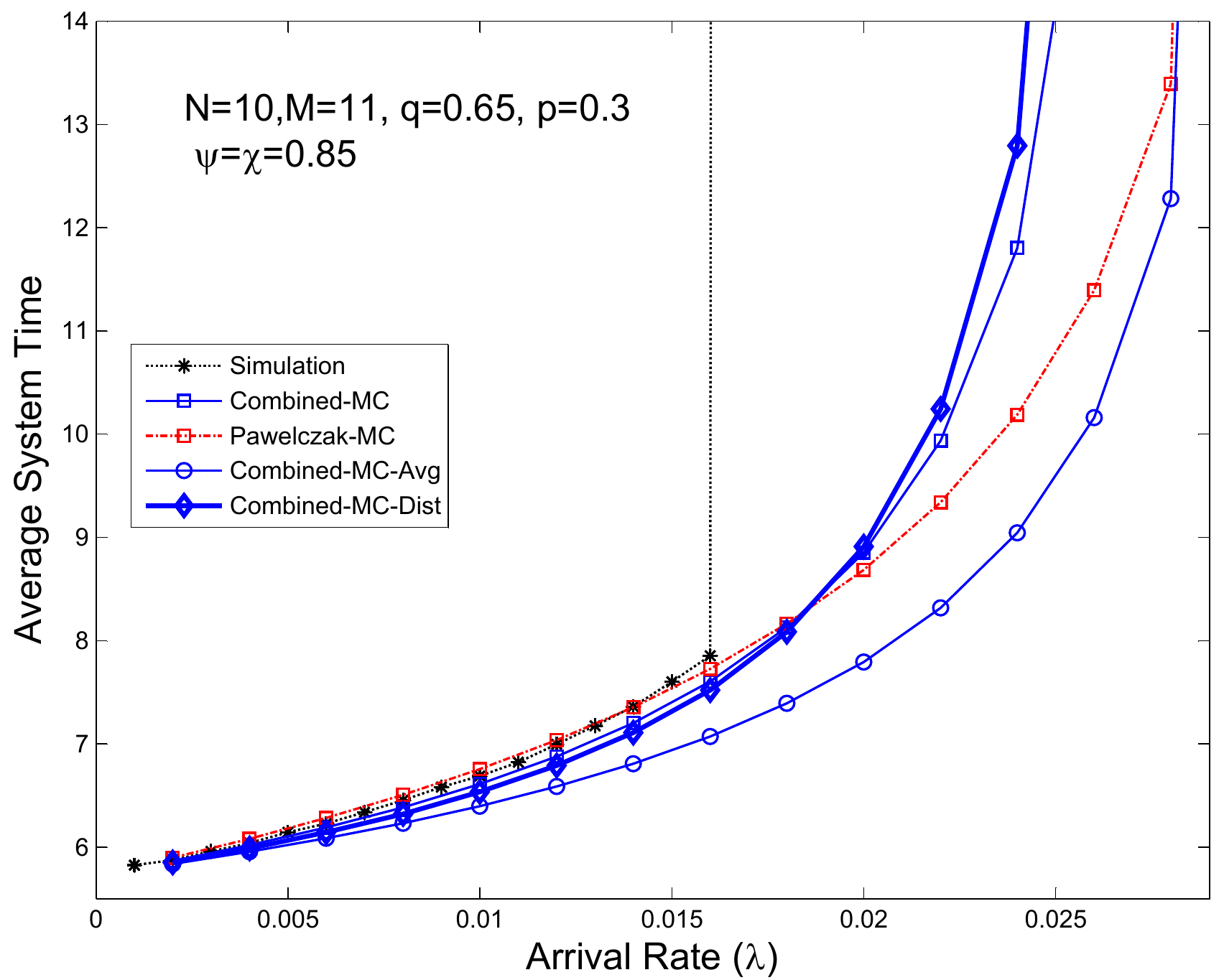}
\caption{Performance comparison of the proposed schemes versus the variation of the arrival rate for the buffering MAC protocol.}
\label{FIg21102013New}%
\vspace{1em}
\includegraphics[width=0.4\columnwidth]{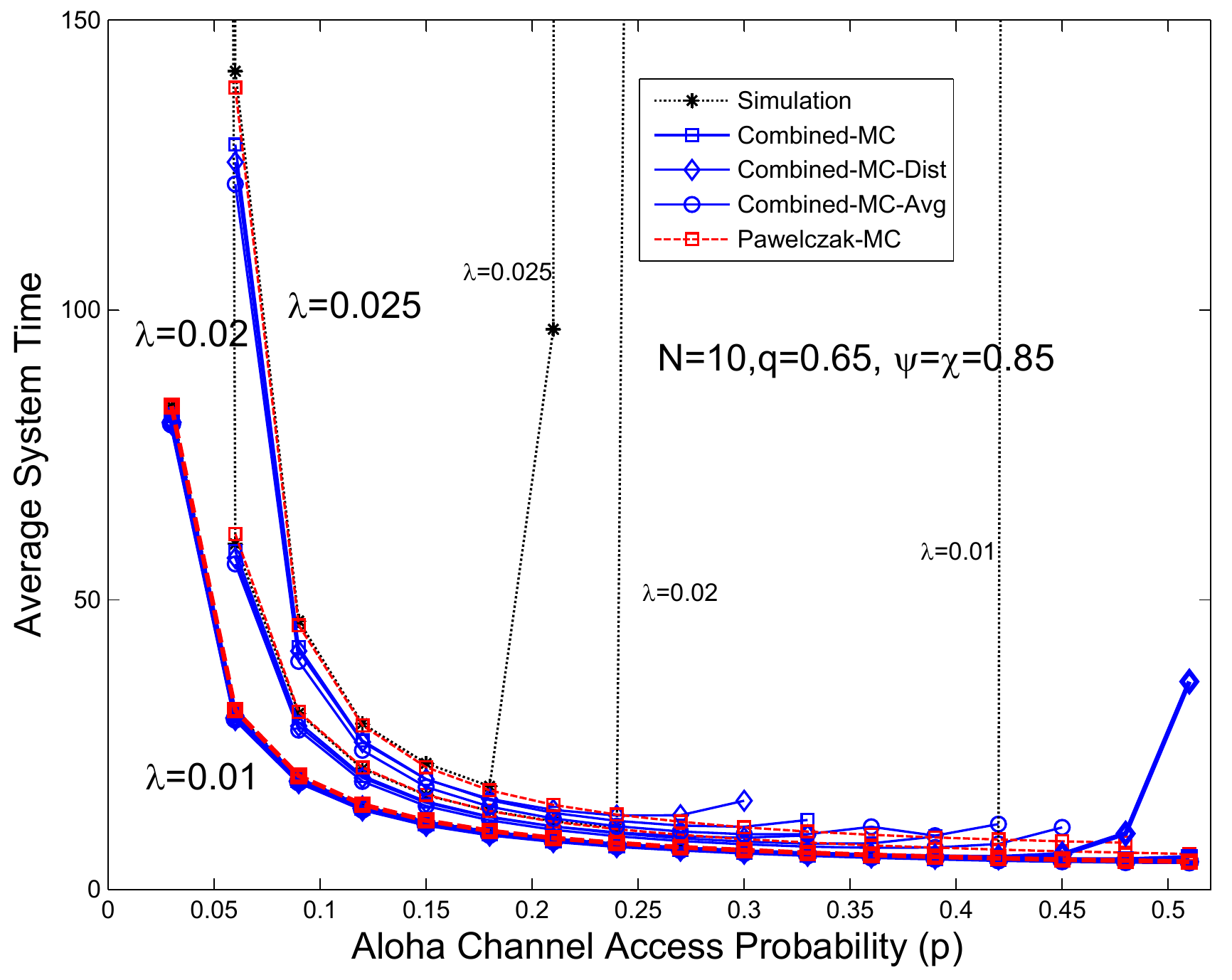}%
\includegraphics[width=0.4\columnwidth]{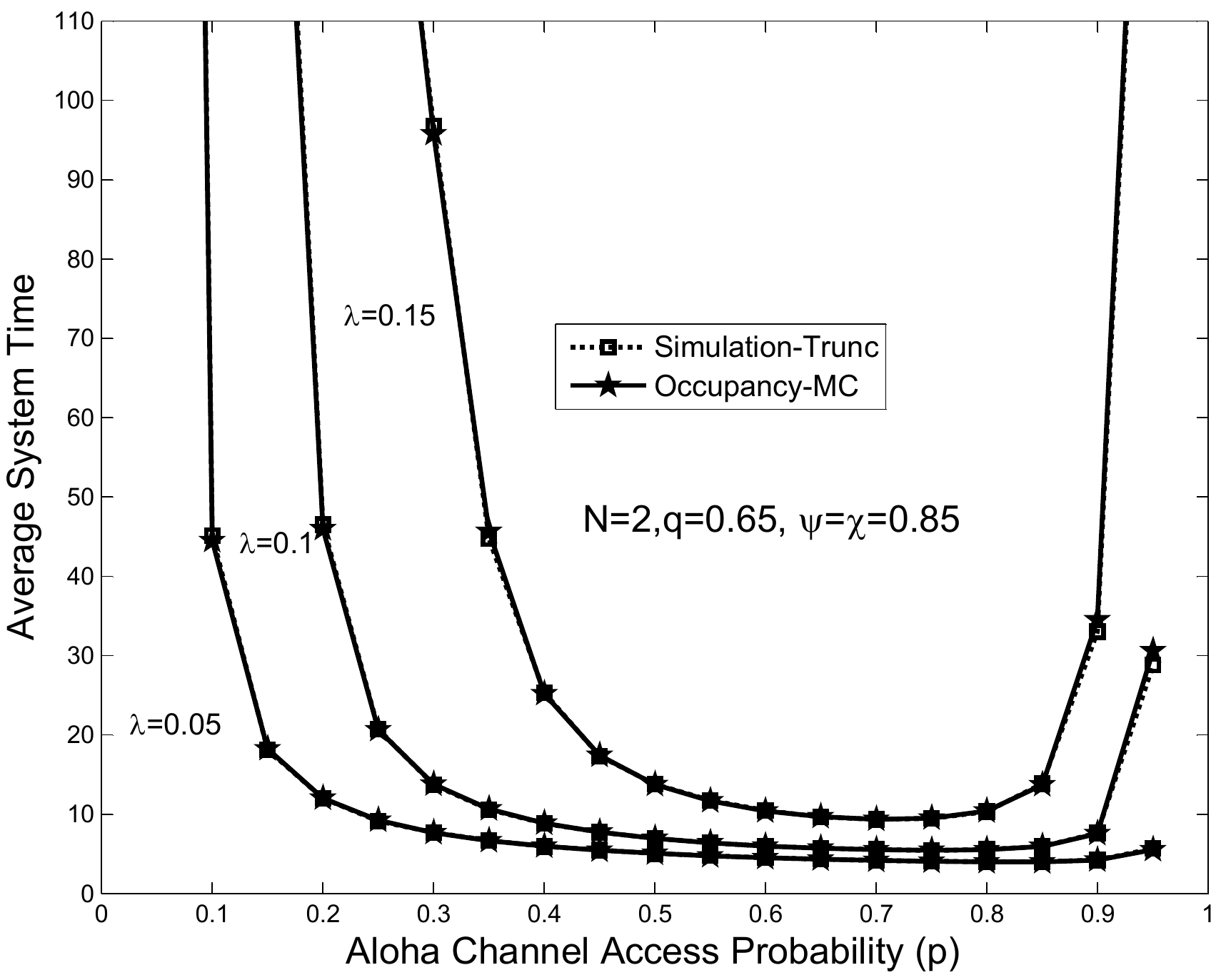}\\%
\includegraphics[width=0.4\columnwidth]{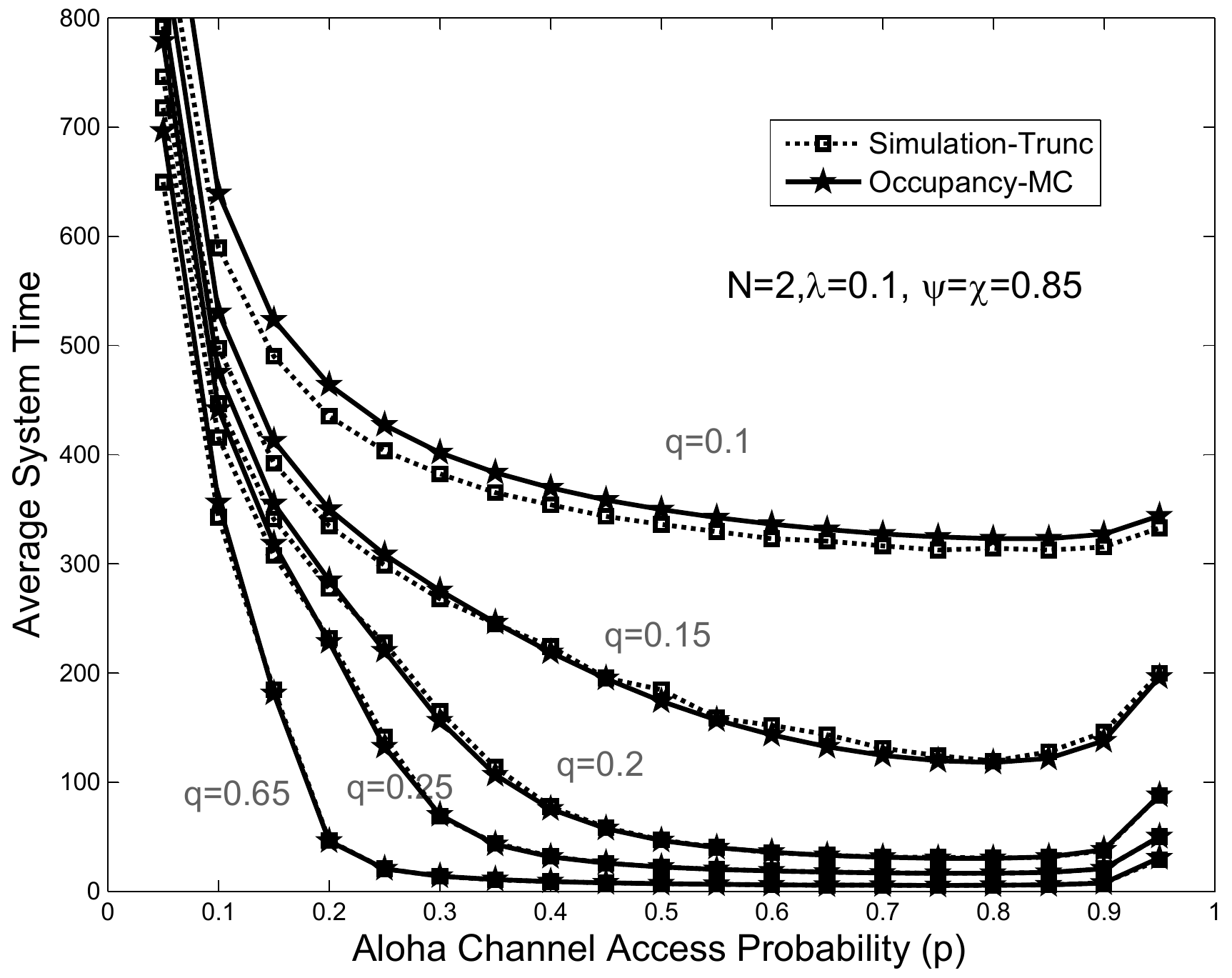}%
\caption{Average system time for the buffering MAC protocol versus the Aloha-type probability of control channel access, for different arrival rate and packet length values.}%
\label{FigPVariable3LambdasOnlySim}%
\end{figure*}

Fig.~\ref{FIg21102013New} shows the average system time as a
function of the arrival rate for the buffering MAC protocol and for different system parameter values. 
It can be first observed that, as expected, the queue occupancy Markov chain 
provides accurate results compared with simulation results, called 'truncated simulation', obtained
with the same limited buffer size assumed to numerically solve the occupancy Markov chain. 
All approximations also provide acceptable results at low arrival rate while for larger values, when
the system gets close to instability, their accuracy decreases. 
For example, for $N=10$ users, the combined-MC approximation is accurate within 3\% for a normal operation region
below 80\% of maximum loading. It was observed through several scenarios that, as expected, the combined Markov chain
provides the best approximation while the average based and distribution based approximations both diverge sooner 
than the two other schemes when the queue approaches the instability region.
In the following, to help the figures readability, we will only present some of the results obtained with the different approximations.

\subsection{Impact of Medium Access Probability}

The probability of medium access on the control channel 
has a major influence on the performance of the network. 
A very small $p$ or a very large $p$ makes the system 
unstable because, in the former case, a packet has to 
wait several slots before the node tries to reserve a channel, 
and, in the latter case, there will be a lot of collisions such that the reservation success probability
will be very low. In Fig.~\ref{FigPVariable3LambdasOnlySim}, 
we can see that for any given scenario of the buffering MAC protocol, there is an optimal value for $p$ which depends
on the arrival rate. 
This optimal value decreases when the total traffic load on the control channel increases due
to either an increase in the arrival rate, an increase in the number of nodes or smaller packets.
Interestingly, it can be observed that by increasing the packet length (smaller $q$), 
even though the total system load and the delay increase because the packets are now longer, 
the optimal probability of channel access $p$ also increases. 
This is due to the fact that with longer packets, the nodes return less frequently to the control channel. There
is thus less competition for channel reservation and a higher probability of control channel access is optimal. 

It is also observed that the optimal control channel access probability is significantly
larger than for the saturated traffic model where the optimal value of $p$ is very small~\cite{pawelczak09}. There is also
a large range of values of $p$ where the performance is similar. However, this range decreases with an increase in traffic load of the system.

\subsection{Impact of Packet Length}

\begin{figure}%
\includegraphics[width=\columnwidth]{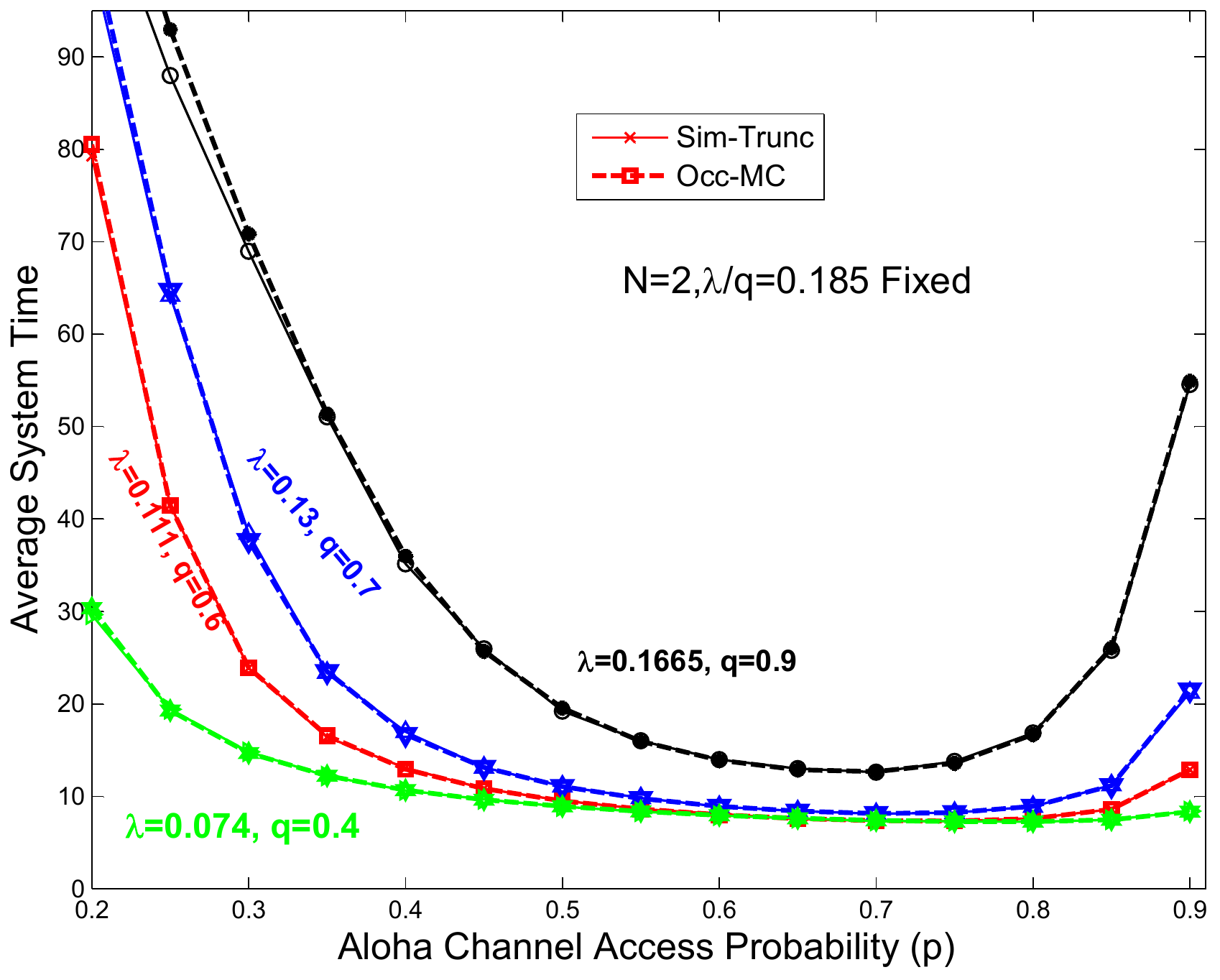}%
\caption{Performance comparison for the buffering MAC protocol for a fixed traffic load of $\lambda/q=0.85$ and $\chi=\psi=0.85$.}%
\label{Figure1116171819102013}%
\end{figure}

As observed in Fig.~\ref{FigPVariable3LambdasOnlySim}, the system time increases when
the system load increases either due to an increase of arrival rate or packet length.
In  Fig.~\ref{Figure1116171819102013}, we investigate the impact of both parameters on the buffering MAC protocol when the traffic load
$\lambda/q$ is fixed. It can be observed that as the packet length increases (and the arrival rate decreases),
the system time decreases. This is due to the fact that in the network model with an Aloha-type medium access algorithm,
the main bottleneck is the competition on the control channel to reserve a data channel. 
For a fixed traffic load, when the packet length increases, 
fewer nodes will therefore participate in the competition and the nodes with a reserved channel 
spend more time on the reserved channel and less time waiting on the control channel to get a reservation success.

\subsection{Impact of Transmission Capacity}
\label{sec:res_tx_cap}

\begin{figure}%
\includegraphics[width=\columnwidth]{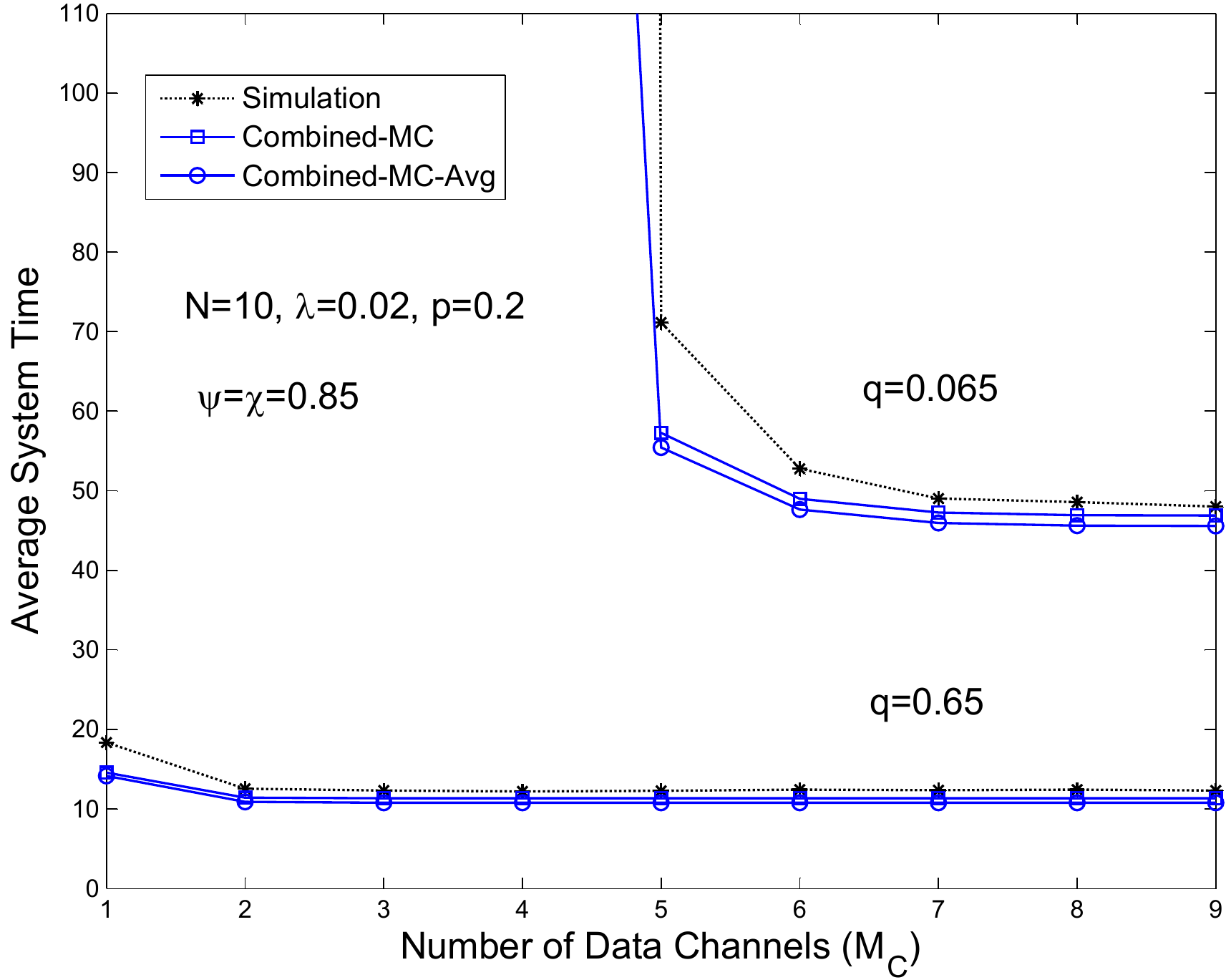}%
\caption{System time as a function of the number of data channels $M_C$ for the buffering MAC protocol.}%
\label{Fig010203112013}%
\end{figure}

We investigate in this section the impact of transmission capacity which depends on the number of available
channels and the channel availability $\psi$.
In Fig.~\ref{Fig010203112013}, the number of nodes is fixed 
to $N=10$ and the number of data channels is increasing 
from $M_C=1$ to $M_C=10$, for two different values of packet length. 
As discussed previously, when the number of nodes is high the probability of success in competition decreases and
all nodes spend a longer time for reserving a channel. Even though there 
are available channels, no node can make a reservation; 
therefore, the increase of the number of channels provides no 
gain for the network. When the packet length is small ($q=0.65$), 
the packet transmission time is short, so a winner again joins quickly 
the competing users and the probability to have even two channels 
busy at the same time is low.  
With a longer packet length ($q=0.065$), fewer users compete and data channels are more utilized. The
system time therefore decreases until 6 data channels are available and marginal gains are achieved
for larger values of $M_C$.

\begin{figure}%
\centering
\includegraphics[width=\columnwidth]{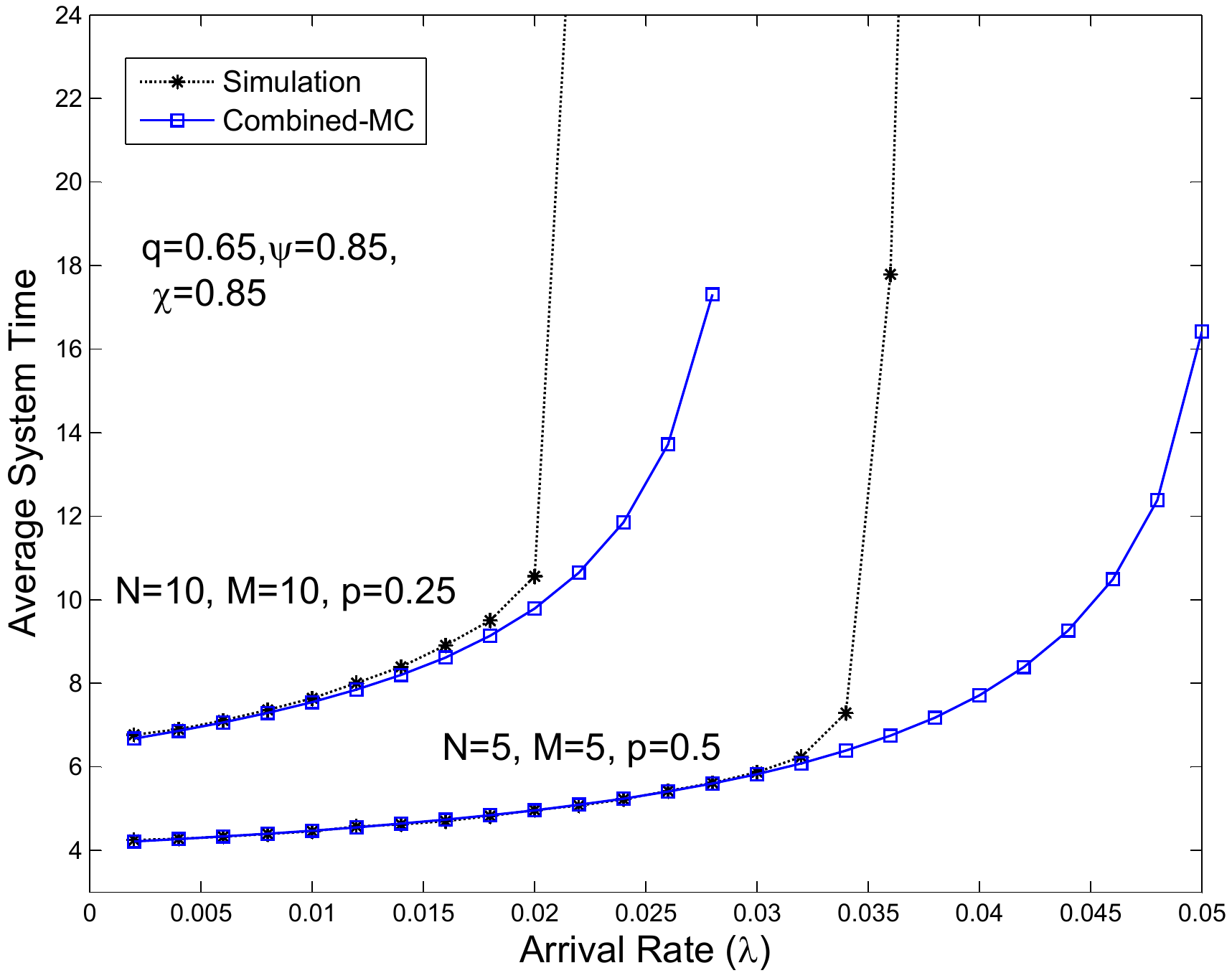}\\
\caption{Performance comparison for the buffering MAC protocol of a network with $M=N=10$ and the clustered network with two clusters of $N=M=5$.}%
\label{Fig25102013NewCorrected2MC3MCNewPNewM2WithSim}%
\end{figure}

The observation that the main bottleneck 
of an Aloha-based cognitive radio MAC protocol is the channel reservation competition on 
the control channel  leads us to propose the concept of channel
and node clustering. That is, instead of having a single control channel
the CR nodes and channels can be divided into clusters where each cluster has its own 
control channel. 
The results presented in Fig.~\ref{Fig25102013NewCorrected2MC3MCNewPNewM2WithSim}
show the performance of such a clustered system.
We assumed a network with 10 nodes and 10 channels and compared the performance when there is a single control channel
and all nodes operate under a single entity of the buffering MAC protocol (i.e., $N=10$ and $M_C=9$) with
the case where the nodes and channels are divided into two clusters and
each cluster has its own control channel and operate under an independent 
entity of the buffering MAC protocol (i.e., $N=5$ and $M_C=4$). The probability of medium access, $p$, has been 
adjusted to the value which minimizes the delay for each case.
It can be observed that even though one additional channel is now used as
a control channel in the two-cluster case, and we thus have fewer data channels, the system time 
has significantly decreased for all values of arrival rate. Furthermore, a larger arrival rate can be 
served when there are two clusters. Those results confirm our hypothesis that the delay bottleneck is due to the reservation
competition on the control channel and the
fact that, depending on the system parameters, it might be more beneficial to add control channels than to add
data transmission capacity. 

\begin{figure}%
\includegraphics[width=\columnwidth]{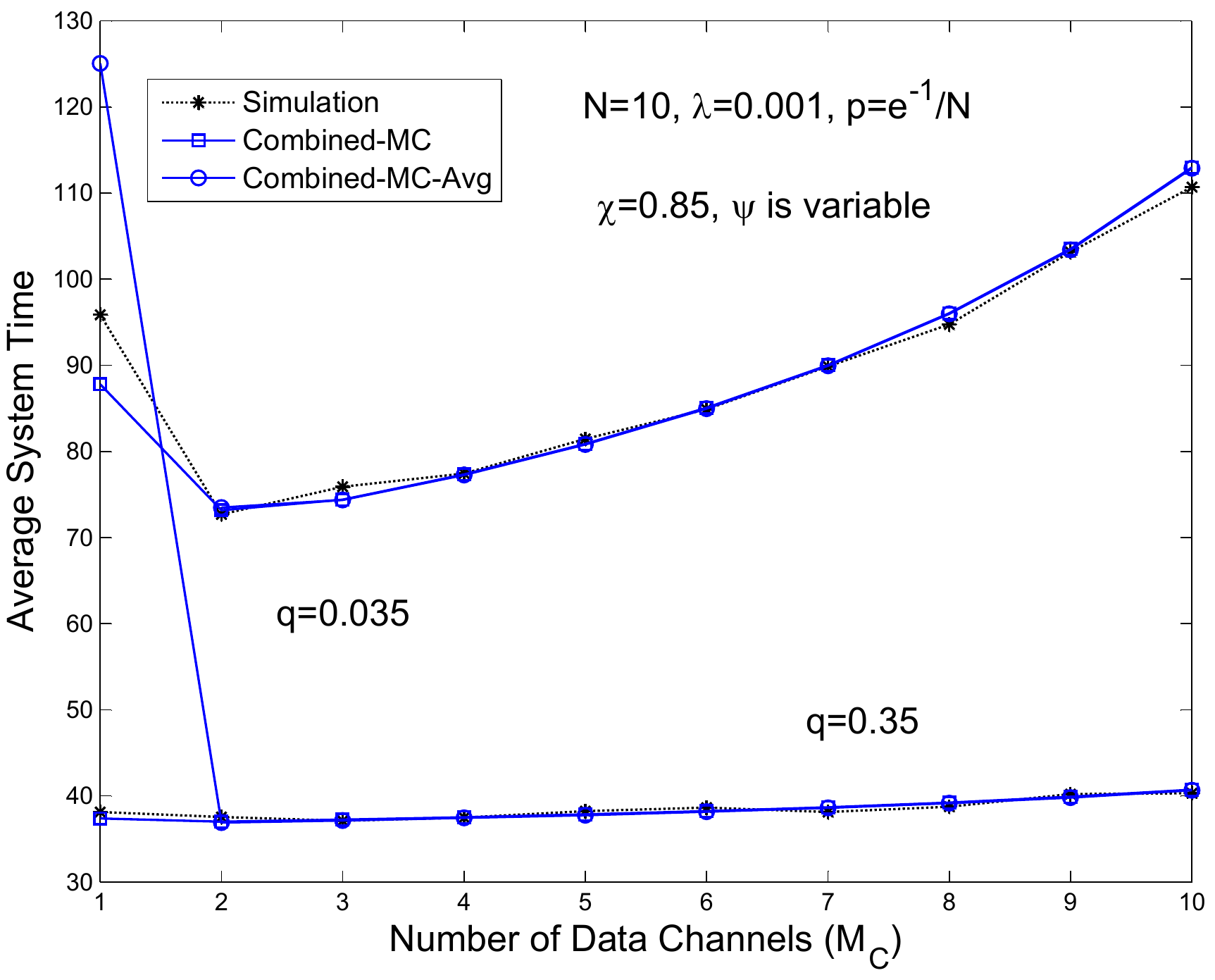}%
\caption{Impact of the variation of the number of data channels $M_C$ when the channel availability, $\psi$, decreases for the buffering MAC protocol. For $M_C=[1\dots 10]$, we have $\psi=[0.85, 0.8, 0.75, 0.7, 0.65, 0.6, 0.55, 0.5, 0.45, 0.4]$.}%
\label{Fig0405112013}%
\end{figure}

Another important issue is the trade-off between the 
number of channels and their availability, when, for instance, 
a cognitive radio network has the option of utilizing a spectrum with a large number of 
channels with a lower availability per timeslot and another spectrum with a 
smaller number of channels with a higher availability.  In Fig.~\ref{Fig0405112013}, we investigate
this tradeoff where the channel availability $\psi$ decreases when $M_C$ increases.  Note that 
we kept the control channel availability constant.
The interesting point to observe is that there is
an optimal point where the average delay is minimized. 
This is due to the fact that because of the control channel access bottleneck, an 
increase in the number of channels does not provide 
any significant gain for the network (see results presented in Fig.~\ref{Fig010203112013}) 
while the decrease in 
availability of the assigned channel increases the transmission time of the packet. 
Such an optimal point exists for any network as a 
function of the system parameters, 
so the analytical queueing results presented in this paper enable the network designer to select the best operating channel set.

\subsection{Switching MAC Protocol}
\label{sec:sim-switch}

\begin{figure}%
\includegraphics[width=\columnwidth]{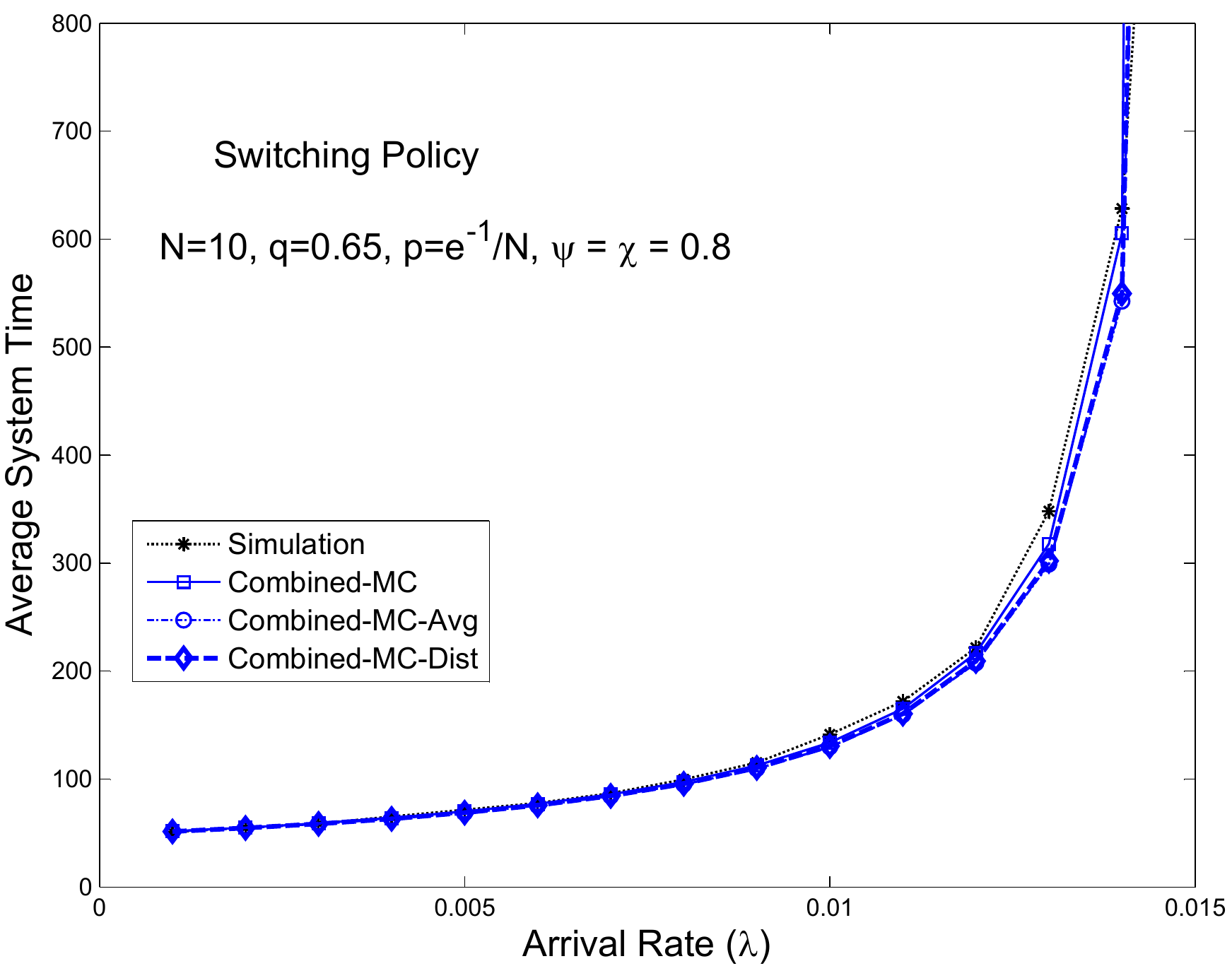}%
\caption{System time for the switching MAC protocol as a function of the arrival rate.}%
\label{Fig29112013}%
\end{figure}

In this section we investigate the performance of the switching MAC protocol. 
Fig.~\ref{Fig29112013} shows the system time as a function of the arrival rate for $N=10$ nodes.
It can first be observed that the analytical approximations provide a good prediction of the 
system performance.
Those results also show that there is a significant performance degradation versus
the results presented previously for the buffering MAC protocol. For similar parameters, we observe approximately $50\%$ lower delay in the buffering case. 
This is due to the fact that for a packet transmission a node has to incur several reservation periods, which
are the delay bottleneck, and furthermore this creates more competition on the control channel, which further
increases the reservation delay.

\begin{figure}%
\centering
\includegraphics[width=0.6\columnwidth]{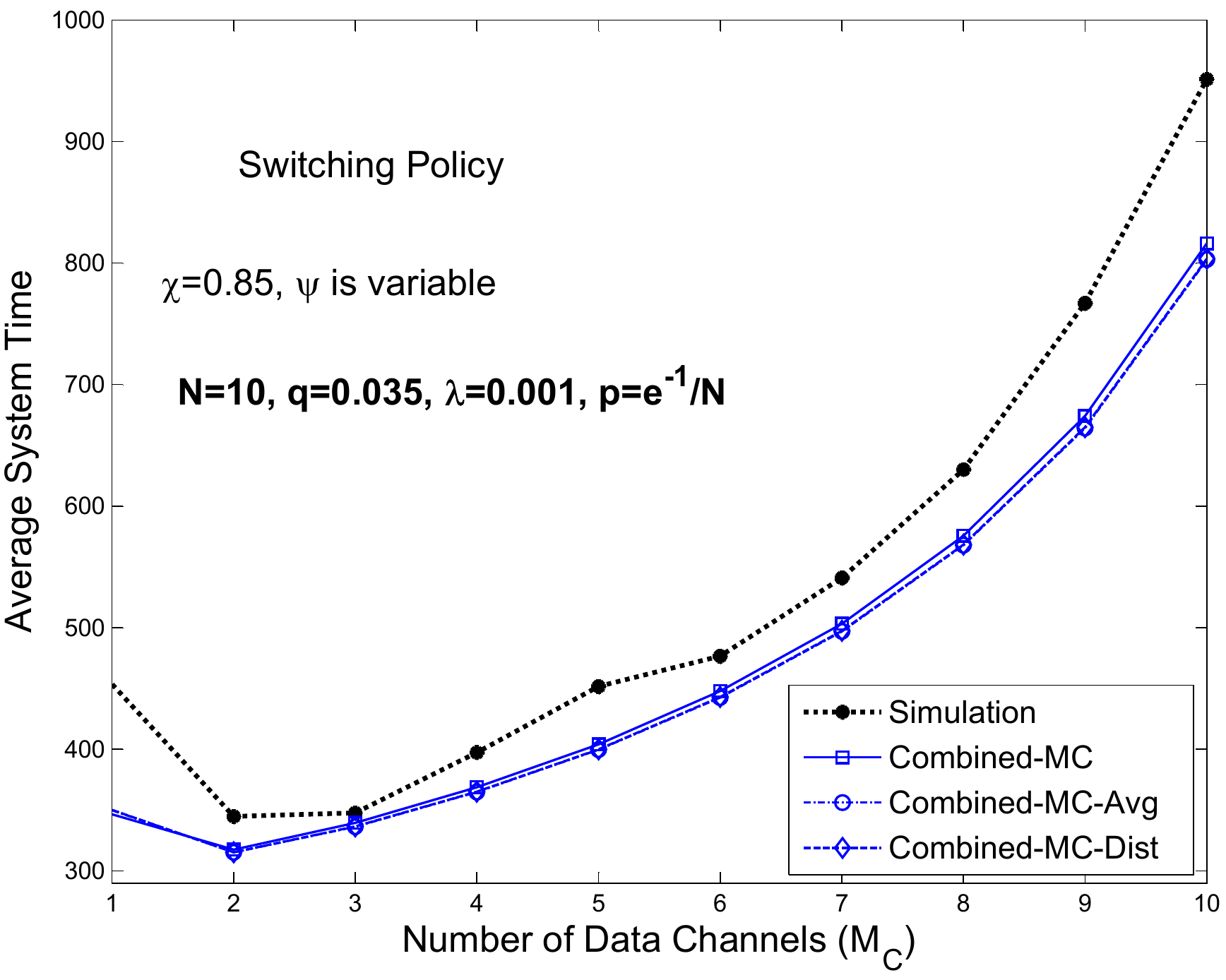}\\
\includegraphics[width=0.6\columnwidth]{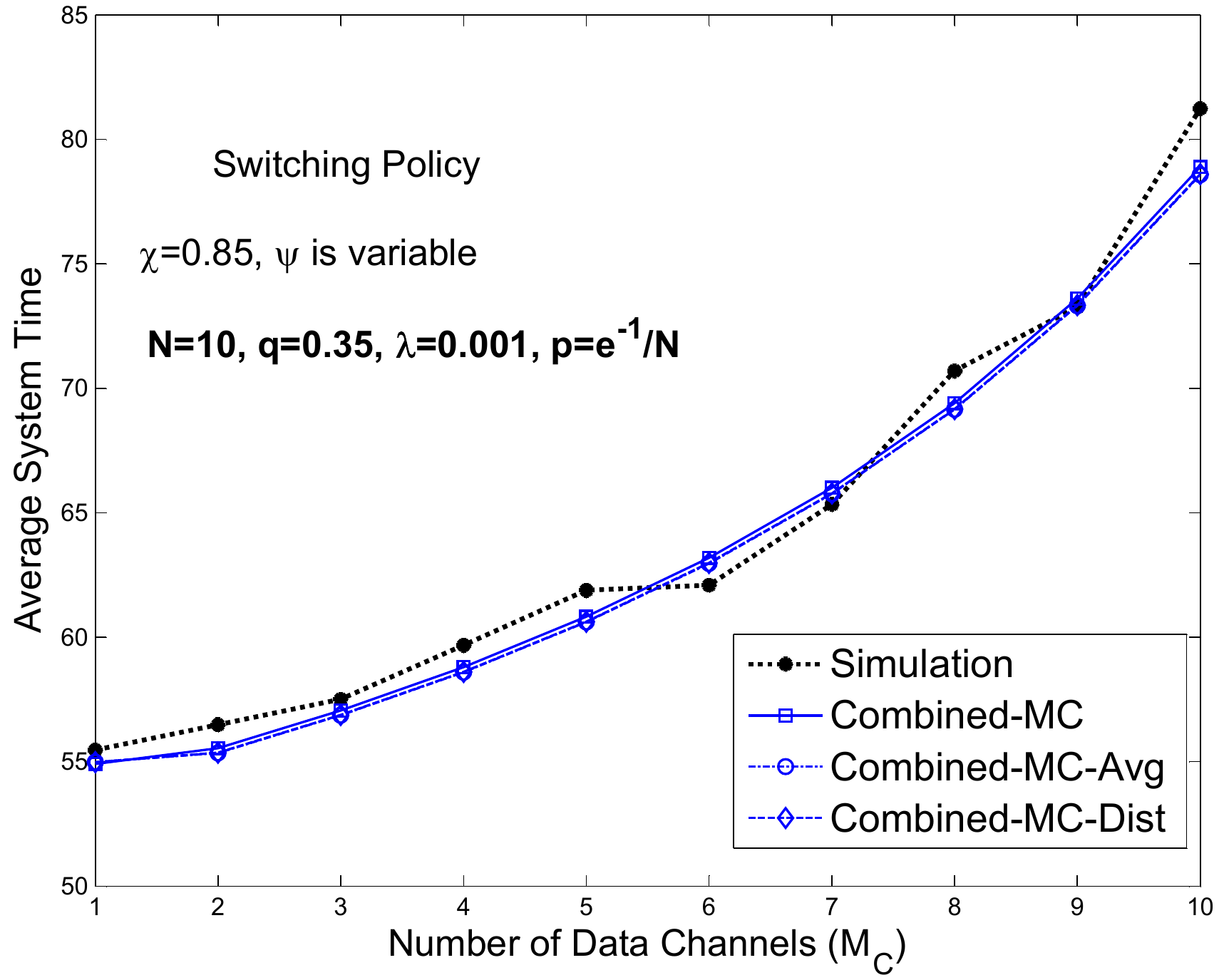}%
\caption{Performance comparison versus the variation of the number of data channels $M_C$ when the channel availability, $\psi$, decreases for the switching MAC protocol.
For $M_C=[1\dots 10]$, we have $\psi=[0.85, 0.8, 0.75, 0.7, 0.65, 0.6, 0.55, 0.5, 0.45, 0.4]$.}
\label{Fig2728112013}
\end{figure}

In Fig.~\ref{Fig2728112013}, the performance versus 
the variation of the number of data channels and channel availability $\psi$
is illustrated. It can easily be 
seen that the performance of the switching policy 
significantly degrades especially when the packet length 
is long. This is due to the fact that the longer the packet are and the more frequent the transmission
occurs, the more often the nodes return to the control channel and wait to get a reservation.
However, we can still observe for large packets $(q=0.035)$ an initial 
improvement when the number of channels increases 
from one to two.

\begin{figure}%
\includegraphics[width=\columnwidth]{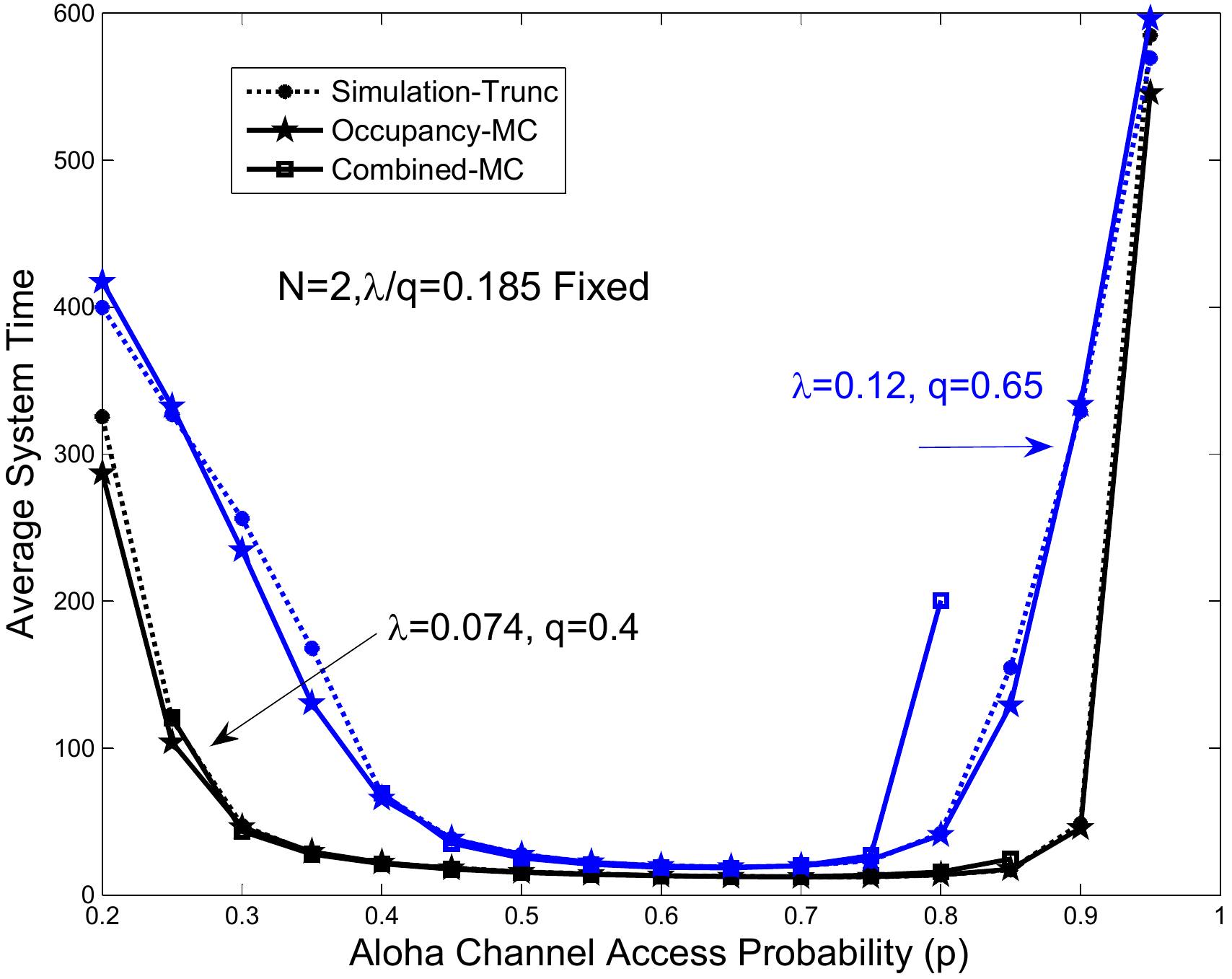}%
\caption{Average system time for the switching MAC protocol versus the Aloha-type probability of control channel access. ($M_C=N, \chi=\psi=0.8$).}%
\label{Fig303132112013.pdf}%
\end{figure}

Fig.~\ref{Fig303132112013.pdf}  shows the impact of Aloha access probability on the packet system time 
for the switching MAC protocol. Those results first
validate the occupancy Markov chain model.
It is also interesting to observe that even though the switching policy 
is more sensitive to an increase in packet length due to the frequent returns to the control channel, 
there are some scenarios, as the one illustrated in the figure, in which having a lower arrival rate with longer packets provides
a better performance than having a higher arrival rate of shorter packets. 



\begin{figure}%
\centering
\includegraphics[width=\columnwidth]{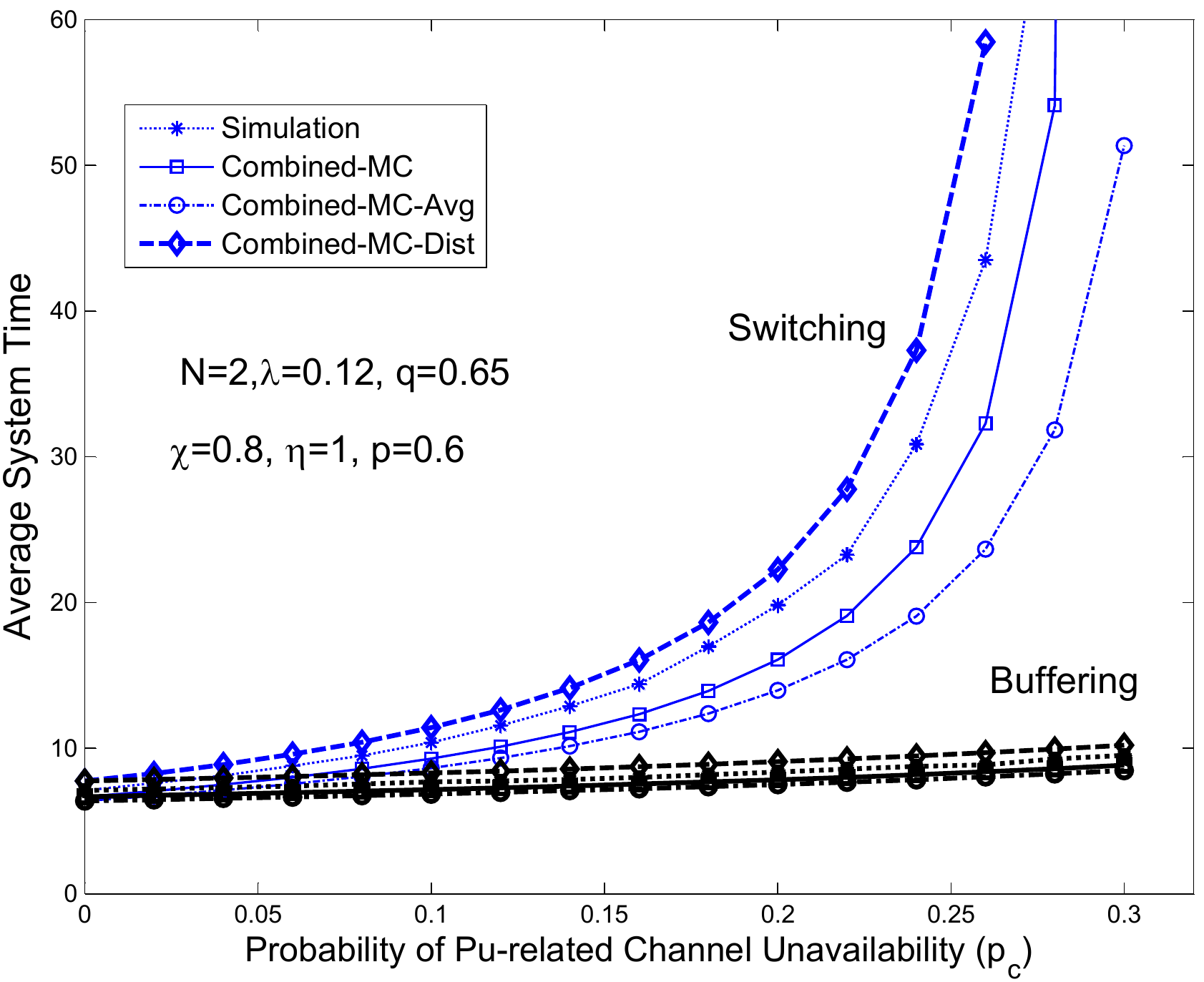}%
\caption{Average system time for the switching and buffering MAC protocols as a function of PU probability of activity ($p_c$).}%
\label{Fig33112013.pdf}%
\end{figure}

Since the switching occurs due to PU arrivals and not fading, 
we investigate the impact of variation of $p_c$ on the system time. The results for the switching and buffering MAC protocols are 
illustrated in Fig.~\ref{Fig33112013.pdf}. When $p_c=0$, the two policies 
are naturally the same. When $p_c$ increases, the performance of the buffering MAC policy
only slightly deteriorates due to a small increase in packet transmission time. However, for the switching
MAC policy, each PU interruption incurs a large penalty of channel reservation. Therefore, the performance
quickly degrades as a function of $p_c$.
It is worth mentioning that we have also ignored the switching time required 
to align the radio in the switching policy \cite{park11}. Naturally, 
assuming a switching time will further degrade the 
performance of the switching policy.

Comparing the results of the buffering and switching 
policies reveals that in all  scenarios, the buffering 
MAC protocol outperforms the switching. 
This is simply due to the homogeneity of the channels and memoryless
PU presence. That is, with the switching policy if the channel is not available,
the node tries to reserve a channel. In the best case, the reservation process
will last one timeslot. However, the reserved
channel has the same probability of being available in the next timeslot as if
the node stayed on the same channel with the buffering policy.
So there is no way for a node to decrease its transmission delay
by switching to another channel.

\section{Conclusion}
\label{sec:conclusion}
In this paper, a delay and queueing analysis for a multi-node network with an 
Aloha-type medium access model was provided. Two different recovery models 
were considered: a waiting and buffering recovery policy where the CR node 
waits for the primary user to vacate the channel and continues the transmission 
on the same channel, and a switching policy where after the appearance of 
primary user, a spectrum handover occurs. It was observed that access to 
the control channel to reserve a data channel is the major bottleneck in Aloha, 
so any approach which decreases the number of competitors, 
such as having fewer but longer packets and node clustering, improves the performance. 
The probability of medium access in Aloha should also be adjusted carefully
to have the minimum delay. With the assumption of having homogenous and 
memoryless channels, a buffering policy always outperforms the switching policy. 
An important area of future work would be to improve the channel occupancy model with a 
Markov chain channel instead of the assumption of independent 
availability per timeslot. Furthermore, since the Aloha access to the control channel represents
a major contributor to the delay, alternative control channel strategies should also be investigated following 
the methodology presented in this paper.

\bibliography{new-refs,telecom}
\bibliographystyle{IEEEtran}

\end{document}